\documentclass[preprintnumbers,article,amsmath,amssymb,floatfix,10pt,prd,onecolumn,
superscriptaddress,nofootinbib]{revtex4-2}
\usepackage{bm}
\usepackage{amsfonts}
\usepackage{latexsym}
\usepackage[latin1]{inputenc}
\usepackage{graphicx}
\usepackage{amsmath}
\usepackage{palatino}
\usepackage{mathpazo}
\usepackage{textcomp}
\usepackage{float}
\usepackage{booktabs}
\usepackage{dcolumn}
\usepackage{ragged2e}
\usepackage{hyperref}
\hypersetup{colorlinks,citecolor=blue}
\hypersetup{colorlinks=true,linkcolor=red,filecolor=magenta,    urlcolor=blue}
\usepackage{amsmath}
\usepackage{xcolor}
\usepackage{orcidlink}
\usepackage{epsfig}
\usepackage[caption=false]{subfig}
\usepackage{commath}

\usepackage{multirow}
\newcommand{\be}{\begin{equation}}
\newcommand{\ee}{\end{equation}}
\newcommand{\bea}{\begin{eqnarray}}
\newcommand{\eea}{\end{eqnarray}}
\newcommand{\eeas}{\end{eqnarray*}}
\newcommand{\beas}{\begin{eqnarray*}}

\def\jnl@style{\it}
\def\aaref@jnl#1{{\jnl@style#1}}

\def\aaref@jnl#1{{\jnl@style#1}}

\def\aj{\aaref@jnl{AJ}}                   
\def\apj{\aaref@jnl{ApJ}}                 
\def\apjl{\aaref@jnl{ApJ}}                
\def\apjs{\aaref@jnl{ApJS}}               
\def\apss{\aaref@jnl{Ap\&SS}}             
\def\aap{\aaref@jnl{A\&A}}                
\def\aapr{\aaref@jnl{A\&A~Rev.}}          
\def\aaps{\aaref@jnl{A\&AS}}              
\def\mnras{\aaref@jnl{Mon.~Not.~Roy.~Astron.~Soc.}} 
\def\prd{\aaref@jnl{Phys.~Rev.~D}}        
\def\prc{\aaref@jnl{Phys.~Rev.~C}}        
\def\prl{\aaref@jnl{Phys.~Rev.~Lett.}}    
\def\qjras{\aaref@jnl{QJRAS}}             
\def\skytel{\aaref@jnl{S\&T}}             
\def\ssr{\aaref@jnl{Space~Sci.~Rev.}}     
\def\zap{\aaref@jnl{ZAp}}                 
\def\nat{\aaref@jnl{Nature}}              
\def\aplett{\aaref@jnl{Astrophys.~Lett.}} 
\def\apspr{\aaref@jnl{Astrophys.~Space~Phys.~Res.}} 
\def\physrep{\aaref@jnl{Phys.~Rep.}}      
\def\physscr{\aaref@jnl{Phys.~Scr}}       
\def\commat{\aaref@jnl{Comm.~Math.~Phys.}} 
\def\science{\aaref@jnl{Science}}          
\def\cqg{\aaref@jnl{Classical Quant.~Grav.}}  
\def\jpcs{\aaref@jnl{JPCS}}                   
\def\ijmpd{\aaref@jnl{Int.~J.~Mod.~Phys.~D}}  
\def\grg{\aaref@jnl{Gen.~Relat.~Gravit.}}     
\def\rpp{\aaref@jnl{Rep.~Prog.~Phys.}}        
\def\npa{\aaref@jnl{Nucl.~Phys.~A}}        
\def\lrr{\aaref@jnl{Living Rev.~Rel.}}     
\def\jcap{\aaref@jnl{J.~Cosmology Astropart.~Phys.}}    
\def\rmp{\aaref@jnl{Rev.~Mod.~Phys.}}   
\def\epjc{\aaref@jnl{Eur.~Phys.~J.~C}} 
\def\plb{\aaref@jnl{~Phy.~Lett.~B}} 
\def\mpla{\aaref@jnl{Mod.~Phy.~Lett.~A}} 
\def\arxiv{\aaref@jnl{arxiv.org}}

\allowdisplaybreaks[1]

\begin{document}

\newpage
\color{black}       
\title{Anisotropic LRS-BI Universe with $f(Q)$ gravity theory}

\author{Pranjal Sarmah\orcidlink{0000-0002-0008-7228}}
\email{p.sarmah97@gmail.com}
\affiliation{Department of Physics, Dibrugarh University, Dibrugarh 786004, 
India}

\author{Avik De\orcidlink{0000-0001-6475-3085}}
\email{avikde@utar.edu.my}
\affiliation{Department of Mathematical and Actuarial Sciences, Universiti 
Tunku Abdul Rahman, Jalan Sungai Long, 43000 Cheras, Malaysia}

\author{Umananda Dev Goswami\orcidlink{0000-0003-0012-7549}}
\email{umananda@dibru.ac.in}
\affiliation{Department of Physics, Dibrugarh University, Dibrugarh 786004, 
India}
%

\begin{abstract}
The possible anisotropic nature in the early phases of the Universe is one of
the interesting aspects of study in cosmology. We investigate the evolution 
of the Universe in terms of few cosmological parameters considering an 
anisotropic locally rotationally symmetric (LRS) Bianchi type-I spacetime 
(LRS-BI) under the $f(Q)$ gravity of symmetric teleparallel theory equivalent 
to the GR (STEGR). In this study we consider two $f(Q)$ gravity models, viz., 
$f(Q) = -\,(Q+2\Lambda)$, a simple model with the cosmological constant 
$\Lambda$ and the power law model, $f(Q) = -\, \alpha Q^n$ with two constant 
parameters $\alpha$ and $n$. Considering a proportionality relation between 
the directional Hubble parameters with a proportionality constant parameter 
$\lambda$ we find a significant contribution of the anisotropic factor in 
the evolution of the early Universe for both models. The power law model 
shows the more dominating effect of anisotropy in comparison to the simple 
model depending on model parameters, especially on the parameter $n$. The 
power law model also shows some possible effects of anisotropy on the 
Universe's evolution in the near future. In addition, both models confirm that 
the anisotropy does not obtain any appreciable signature in the current stage 
of the Universe. From our analysis we also set rough constraints on our model 
parameters as $0.5 \le \lambda \le 1.25$, $0.75 \le \alpha \le 1.5$ and 
$0.95 \le n \le 1.05$.

\end{abstract}

\maketitle

\section{Introduction}\label{sec1}

The cosmological principle suggests that the perspective of an observer on the cosmos is independent of both his position and the direction in which he is looking. This very useful assumption implies that the present-day Universe is largely isotropic and homogenous, and can be modelled perfectly by a Friedmann-Lema\^itre-Robertson-Walker (FLRW) spacetime metric. However, this may not have been the case in the very early stage of the Universe or in the distant future. Recent results from the Wilkinson Microwave Anisotropy Probe (WMAP) \cite{wmap,wmap1,wmap2} show that the classic isotropic and homogeneous model of the Universe requires some tweaking to account for the observational evidence. The remarkable aspect of the inflationary paradigm \cite{udg} that has been supported by observation is that it isotropizes the early Universe into the FLRW geometry we observe today. One must still allow for the possibility of spatial inhomogeneity and anisotropy, and then examine its development into the observed quantity of homogeneity and isotropy, for a complete version.
{However, the observational evidences of smallness of Cosmic Microwave
Background (CMB) quadrupole \cite{cornish_2004}, low anomalies in the large 
scale angular distribution of CMB power spectrum \cite{costa_2004} etc.\ 
suggest some sort of symmetries in the Universe.} {Again, studies show 
that a plane symmetric large scale geometry of the Universe with the 
eccentricity of the order $10^{-2}$ can bring the quadrupole amplitude match 
to the observational data without affecting the higher order multipoles of the 
temperature anisotropy in the CMB angular power spectrum \cite{Jaffe3}. 
Further, the polarization analysis of electromagnetic radiation propagating 
over large cosmological distances also indicates the symmetry axis in 
the large scale geometry of the Universe \cite{akarsu_2010}.} {To 
address the symmetric nature of the Universe,} 
one option is the locally symmetric Kantowski-Sachs spacetime, but one may also start with the Bianchi type cosmological models, which together make up a sizable and nearly exhaustive class of relativistic cosmology models that are homogeneous but not necessarily isotropic. Furthermore, the isotropic ones, which may be seen as a specific sub-case \cite{pitrou}, can be properly understood by examining a model that is almost like FLRW but has fewer symmetries. {The 
simplest Bianchi model which is very suitable for addressing the Universe with 
such a symmetry axis is the locally rotationally symmetric (LRS) Bianchi 
type-I (BI) model. In LRS-BI model, the metric has the spatial section with 
planner symmetry and an axis of symmetry \cite{Tedesco_2018}. Thus it has two 
equivalent longitudinal directions and a transverse direction. The metric of 
it in Cartesian coordinates is}

\begin{equation} \label{metric}
ds^2 = -dt^2 + A^2(t) dx^2 + B^2(t) (dy^2 + dz^2).
\end{equation} 
{Here, the longitudinal directions are $y$ and $z$ and the transverse 
direction is along the $x$-axis. The metric describes an ellipsoidal expansion 
of space with the cosmic time which supports the CMB observations and hence it 
is widely used \cite{Tedesco_2018}. Also} the above metric has a unique 
feature in the sense
that starting from the most general linearly perturbed spatially flat FLRW
metric, under synchronous gauge and the homogeneity bound, it is possible to
obtain such an anisotropic metric (\ref{metric}) \cite{jcap/AD}. Furthermore,
Bianchi cosmologies have gained popularity in observational cosmology in
recent decades, as WMAP data \cite{wmap} suggest that the standard
cosmological model with a positive cosmological constant resembles the
Bianchi morphology \cite{Jaffe,Jaffe1,Jaffe2,Jaffe3,Jaffe4}. Furthermore,
 according to these findings, the Universe should have been able to develop a
somewhat anisotropic spatial geometry in spite of the inflation, which runs
 counter to the predictions of standard inflationary theories
  \cite{Guth,Guth1,Guth2,Guth3,Guth4,Guth5,Guth6}. Very recently, a wide range 
of Bianchi cosmology with the observational data have been studied (see 
details in \cite{espo,A1,A2,A3,A4,A5}). {Investigations on various 
aspects of 
anisotropic cosmology in classical GR by using the LRS-BI metric have been 
done extensively so far. In Ref.~\cite{Tedesco_2018} the connection between 
deceleration parameter and cosmic shear, and jerk parameter and ellipsoidal 
Universe has been studied in GR theory by using the LRS-BI metric. Another 
work containing the study of the ellipsoidal Universe in Braneworld cosmology 
has been found in Ref.~\cite{Ge_2007}. L.~Campanellie et al.~in 2011 had 
tested the isotropy of the Universe with type Ia supernovae data in the LRS-BI 
framework \cite{Camp_2011}. The work on LRS-BI model to examine the anisotropy 
in the Universe by using the concept of dynamically anisotropic dark energy 
and constant deceleration parameter for perfect fluid is found in 
Ref.~\cite{akarsu_2010}. A wide range of studies have also been carried out  
in Modified Theories of Gravity (MTGs) like $f(R)$, $f(R,T)$
etc.~by using the LRS-BI model of the Universe. X.~Liu et al.~in 2017 had 
studied the cosmological dynamics of an uniform magnetic field in the LRS-BI 
Universe for the viable $f(R)$ models \cite{Liu_2017}. The study of 
anisotropic cosmological model in $f(R,T)$ gravity with variable deceleration 
parameter for the LRS-BI metric has been found in Ref.~\cite{Sahoo_2017}. 
Moreover, the study of de-Sitter and bounce solution in $f(R,T)$ cosmology 
for LRS-BI Universe has been carried out in the Ref.~\cite{mishra_2019}. 
Recently the works on anisotropy cosmology by using the LRS-BI metric have 
been also performed in the $f(Q)$-gravity theory. Various cosmological 
profiles, like 
energy density, equation of state, skewness parameters have been studied in 
$f(Q)$-gravity by using LRS-BI metric in Ref.~\cite{deepjc}. Another work on 
the anisotropic Universe in $f(Q)$-gravity with the consideration of Hybrid 
expansion law (HEL) for the average scale factors has been accomplished in 
Ref.~\cite{Devi_2022}.}

In addition, the accelerated expansion of the Universe was confirmed in 1998 by the Supernovae Cosmology Project, which utilised IA Supernovae data \cite{riess}. This gives rise to a variety of explanations for the acceleration. In general relativity (GR), the existence of a yet undetected unknown kind of energy in the Universe, known as dark energy (DE), with exotic properties such as negative pressure results in a negative equation of state (EoS) parameter. To circumvent the mysterious DE, as an alternative to GR, modified gravity theories were investigated by altering the Einstein-Hilbert action while maintaining the geometry. $f(R)$-gravity theories are the simplest and most successful in this regard. Other types of theories, such as the teleparallel theory equivalent to the GR (TEGR) and the symmetric teleparallel theory equivalent to the GR (STEGR), have been addressed in the past by modifying the underlying geometry without disrupting the Lagrangian.
In these theories, flat space is considered, and the extremely special (symmetric and metric-compatible) Levi-Civita connection used in GR is replaced by an affine connection with either non-vanishing torsion (in TEGR) or non-metricity (in STEGR) as the guiding force of gravity, and its extension $f(Q)$ theory ($Q$ is the non-metricity scalar) was proposed in \cite{Jimenez/2018}. 

Recently, the $f(Q)$-gravity becomes the point of wide attention, see 
\cite{accfQ1,accfQ2,fQfT,fQfT1,fQfT2,fQfT3,deepjc,zhao,gde,lin,cosmography,signa,redshift,perturb,dynamical1,signa,prd/AD,Q1,Q2,Q3,Q4,Q5,Q6} and the references therein for the recent theoretical studies and cosmological and astrophysical applications. However, the majority of the aforementioned cosmological studies have focused on exploring the present interests of the Universe by considering the spatially flat isotropic and homogeneous FLRW metric in Cartesian coordinates. To also note that, the particular line element, so chosen, is automatically a coincident gauge, which reduces the covariant derivative to merely a partial derivative and simplifies the calculation considerably. In our present work we are aiming to study the evolution of the Universe in $f(Q)$-gravity starting from an anisotropic but homogeneous background metric which offers a broader outlook in this new gravity theory and its cosmological application. 

The present article is organized as follows: In the next section the basic
formalism of $f(Q)$-gravity has been discussed. The equations of motion for
the LRS-BI model defined by the metric \eqref{metric} 
in $f(Q)$-gravity are derived in section \ref{sec3}. In section 
\ref{sec4} the expressions of DE EoS and other cosmological parameters are 
developed for the LRS-BI model. For two different $f(Q)$ models we analyse 
the cosmological parameters using the numerical calculations and corresponding 
graphical representations in section \ref{sec5}. Finally, in section 
\ref{sec6} summary and conclusion of the work are drawn with the future 
prospects of the study.      

\section{Basic Formalism of $f(Q)$ Gravity}
\label{sec2}
In $f(Q)$ gravity theory, the spacetime is constructed by using the symmetric 
teleparallelism and non-metricity condition, that is $R^\rho{}_{\sigma\mu\nu} = 0$ and $Q_{\lambda\mu\nu} := \nabla_\lambda g_{\mu\nu} \neq 0$. The associated connection coefficient is given by 
\begin{equation} \label{connc}
\Gamma^\lambda{}_{\mu\nu} = \mathring{\Gamma}^\lambda{}_{\mu\nu} +L^\lambda{}_{\mu\nu}, 
\end{equation}
where $\mathring{\Gamma}^\lambda{}_{\mu\nu}$ is the Levi-Civita connection 
and $L^\lambda{}_{\mu\nu}$ is the disformation tensor. The disformation tensor 
is expressed as  
\begin{equation*}
L^\lambda{}_{\mu\nu} = \frac{1}{2} (Q^\lambda{}_{\mu\nu} - Q_\mu{}^\lambda{}_\nu - Q_\nu{}^\lambda{}_\mu) \,.
\end{equation*}
In addition, we define the superpotential tensor
\begin{equation} \label{P}
P^\lambda{}_{\mu\nu} := \frac{1}{4} \left( -2 L^\lambda{}_{\mu\nu} + Q^\lambda g_{\mu\nu} - \tilde{Q}^\lambda g_{\mu\nu} -\frac{1}{2} \delta^\lambda_\mu Q_{\nu} - \frac{1}{2} \delta^\lambda_\nu Q_{\mu} \right),
\end{equation}
and using it the non-metricity scalar is defined as 
\begin{equation} \label{Q}
Q = -Q_{\lambda\mu\nu}P^{\lambda\mu\nu}. 
\end{equation}
The action of $f(Q)$ gravity is given by
\begin{equation*}
S = \int \left[\frac{1}{2\,\kappa}\,f(Q) + \mathcal{L}_m \right] \sqrt{-g}\,d^4 x,
\end{equation*}
where $\kappa = 8\pi G_N$, $G_N$ being the usual Newtonian gravitational 
constant, $g$ is the determinant of the metric tensor and $\mathcal{L}_m$ is 
the matter Lagrangian. By varying the action with respect to the metric, we 
obtain the field equations of $f(Q)$ gravity as 
\begin{equation} \label{FE1}
\frac{2}{\sqrt{-g}} \nabla_\lambda (\sqrt{-g}\,f_Q\,P^\lambda{}_{\mu\nu}) +\frac{1}{2}\,f(Q)\, g_{\mu\nu} + f_Q(P_{\nu\rho\sigma}\, Q_\mu{}^{\rho\sigma} -2P_{\rho\sigma\mu}\,Q^{\rho\sigma}{}_\nu) = -\kappa\, T_{\mu\nu},
\end{equation}
where $f_Q$ represents the derivative of $f(Q)$ with respect to non-metricity 
scalar $Q$ and $T^m_{\mu\nu}$ is the energy-momentum tensor of the matter 
field.  

On the other hand, using the connection coefficient (\ref{connc}), we can 
have the following relations between the curvature tensors corresponding 
to $\Gamma$ and $\mathring{\Gamma}$:
\begin{equation}
R^\rho{}_{\sigma\mu\nu} = \mathring{R}^\rho{}_{\sigma\mu\nu} + \mathring{\nabla}_\mu L^\rho{}_{\nu\sigma} - \mathring{\nabla}_\nu L^\rho{}_{\mu\sigma} + L^\rho{}_{\mu\lambda}L^\lambda{}_{\nu\sigma} - L^\rho{}_{\nu\lambda} L^\lambda{}_{\mu\sigma}
\end{equation}
and so
\begin{align*}
R_{\sigma\nu} &= \mathring{R}_{\sigma\nu} + \frac{1}{2} \mathring{\nabla}_\nu  Q_\sigma + \mathring{\nabla}_\rho L^\rho{}_{\nu\sigma} -\frac{1}{2} Q_\lambda L^\lambda{}_{\nu\sigma} - L^\rho{}_{\nu\lambda}L^\lambda{}_{\rho\sigma}, \nonumber \\[5pt]
R &= \mathring{R} + \mathring{\nabla}_\lambda Q^\lambda - \mathring{\nabla}_\lambda \tilde{Q}^\lambda -\frac{1}{4}Q_\lambda Q^\lambda +\frac{1}{2} Q_\lambda \tilde{Q}^\lambda - L_{\rho\nu\lambda}L^{\lambda\rho\nu} \,.
\end{align*}
Therefore, by using the symmetric teleparallelism condition, we can rewrite 
the field equations in (\ref{FE1}) as
\begin{equation} \label{FE2}
f_Q \mathring{G}_{\mu\nu} + \frac{1}{2}\,g_{\mu\nu}(f(Q)-Qf_Q) + 2f_{QQ} \mathring{\nabla}_\lambda Q P^\lambda{}_{\mu\nu} = -\kappa\, T_{\mu\nu},
\end{equation}
where $$\mathring{G}_{\mu\nu} = \mathring{R}_{\mu\nu} - \frac{1}{2} g_{\mu\nu} \mathring{R}$$ is the Einstein tensor. {Being a metric-affine theory, 
the connection acts as a dynamic variable in $f(Q)$ theory, however it has 
been shown in Refs.~\cite{dl, jim} that in the coincident gauge choice adopted 
in the present setting, the affine connection's field equation is trivially 
satisfied in a model-independent manner. Hence, in this article our sole 
attention is devoted to the metric field equation \eqref{FE2}.}

\section{Equations of motion in the LRS-BI model}\label{sec3}

For the current scenario, the LRS-BI metric in Cartesian coordinates given by 
equation \eqref{metric} is also a coincident gauge. We have the directional 
Hubble parameters $H_x=\dot{A}/{A}$, $H_y=\dot{B}/{B}$ and the average Hubble 
parameter 
\begin{align}
(H_x+2H_y)/3=H=\dot{a}/{a},
\end{align} 
where $a$ is considered to be the 
average scale factor in the discussed anisotropic Universe. The anisotropic 
evolution of the Universe is characterised by the shear scalar $\sigma$ 
defined by 
\begin{equation}\label{sigma}
\sigma^2=\frac{1}{3}(H_x-H_y)^2=3(H-H_y)^2.   
\end{equation}
We can calculate the non-metricity scalar $Q$ in this spacetime metric as 
\begin{equation}\label{Q}
Q=2H_y^2+4H_xH_y.
\end{equation}
From the field equations \eqref{FE2}, we derive the following equations of 
motion:
\begin{equation}\label{eom1}
\kappa \rho = \frac{f(Q)}{2} - 2f_Q \left[2 H_xH_y + H_y^2 \right], 
\end{equation}
\begin{equation} \label{eom2}
\kappa p_x = -\frac{f(Q)}{2} +2 f_Q \left[ 3HH_y + \dot{H}_y \right] +2H_y \dot{Q} f_{QQ}, 
\end{equation}
\begin{equation} \label{eom3}
\kappa p_y = -\frac{f(Q)}{2} + f_Q \left[ 3 H(3H-H_y) + \dot{H}_x+\dot{H}_y \right] + \left( H_x +H_y \right) \dot{Q} f_{QQ}, 
\end{equation}
where $\rho$ is the density of the matter field, and $p_x$ and $p_y$ are
the pressures of the field along $x$ and $y$ directions respectively. 
Under a situation of isotropic pressure, $p_x=p_y=p$ we have 
\begin{equation}\label{fh}
f_Q(H-H_y)=k\,a^{-3}, 
\end{equation}
for a constant $k$. Therefore, using equation (\ref{sigma}), alternately 
equation (\ref{Q}) can be expressed as 
\begin{equation}\label{Qnew}
Q=6H^2-\frac{6k^2}{f^2_Q \, a^6}=6H^2-2\sigma^2.
\end{equation}
This is a handy expression to calculate the non-metricity scalar once we have
the expressions of the average Hubble parameter and the shear scalar of the
considered anisotropic Universe.

\section{Dark energy EoS and other cosmological parameters}\label{sec4}
The field equations (\ref{FE2}) can be equivalently written in the following 
effective form:
\begin{equation}
\mathring{G}_{\mu\nu}=-\frac{\kappa}{f_Q}T^{eff}_{\mu\nu}=-\frac{\kappa}{f_Q} T_{\mu\nu}+T^{DE}_{\mu\nu},
\end{equation} 
where $-\kappa/f_Q$ is the effective gravitational constant and for its 
positivity in our construction, we assume $f_Q<0$. The DE component 
emerged from the modification of STEGR and is given by
\begin{equation}
T^{DE}_{\mu\nu}=-\frac{1}{f_Q}\left[\frac{1}{2}\,g_{\mu\nu}(f(Q)-Qf_Q) + 2f_{QQ} \mathring{\nabla}_\lambda Q P^\lambda{}_{\mu\nu}\right].
\end{equation}
Moreover, the Friedmann like equations in the STEGR determine the cosmology 
of an anisotropic Universe given by the metric (\ref{metric}) and can be 
calculated from  equations (\ref{eom1}) and (\ref{eom2}) as (for simplicity 
we assume $\kappa=1$) 
\begin{eqnarray}\label{rhoeff}
\rho^{eff}&=&\rho-\frac{f(Q)}{2}+3H^2f_Q-\frac{3\,k^2}{f_Q \, a^6}
\notag\\&=&\rho-\frac{f(Q)}{2}+3H^2f_Q-\sigma^2 f_Q,
\end{eqnarray}
and
\begin{equation}\label{peff}
p^{eff}=p+\frac{f(Q)}{2}-3H^2f_Q-2H\dot{Q}f_{QQ}+\frac{2k\dot{Q}f_{QQ}}{f_Q\, a^3}.
\end{equation}
The additional terms arising in these two equations are due to the 
non-metricity of spacetime which behave like a fictitious DE fluid. 
Thus these DE fluid components evolving due to the non-metricity can 
be given by
\begin{equation}\label{rhode}
\rho^{DE}=-\frac{f(Q)}{2}+3H^2f_Q-\frac{3k^2}{f_Q \, a^6},
\end{equation}
\begin{equation}\label{pde}
p^{DE}=\frac{f(Q)}{2}-3H^2f_Q-2H\dot{Q}f_{QQ}+\frac{2k\dot{Q}f_{QQ}}{f_Q a^3}.
\end{equation}
On a comparison with the spatially flat FLRW case \cite{prd/AD}, we could 
identify two novel terms $-3k^2/f_Q a^6$ and $2k\dot{Q}f_{QQ}/f_Q a^3$ 
appearing in the energy and pressure equations, respectively. These 
terms are arised from the anisotropy of the considered spacetime and these 
can be noted as the $\rho_\sigma$ and $p_\sigma$, respectively.

Now the effective EoS can be written by using equations (\ref{rhoeff}) and 
(\ref{peff}) as follows:
\begin{equation}\label{eoseff}
\omega^{eff} := \frac{p^{eff}}{\rho^{eff}}
= \frac{p+\frac{f(Q)}{2}-3H^2f_Q-2H\dot{Q}f_{QQ}+\frac{2k\dot{Q}f_{QQ}}{f_Q a^3}}{\rho-\frac{f(Q)}{2}+3H^2f_Q-\sigma^2 f_Q}
\end{equation}
Further, the DE EoS can also be written by using equations (\ref{rhode}) 
and (\ref{pde}) as given by
\begin{equation}\label{eosde}
\omega^{DE} := \frac{p^{DE}}{\rho^{DE}}
= -\frac{\frac{f(Q)}{2}-3H^2f_Q-2H\dot{Q}f_{QQ}+\frac{2k\dot{Q}f_{QQ}}{f_Q a^3}}{\frac{f(Q)}{2}-3H^2f_Q+\sigma^2 f_Q}.
\end{equation}

The effective EoS and DE EoS will help us to understand the different 
stages of the evolution of the Universe. Further, we have derived the 
expression of  Hubble parameter in $f(Q)$ gravity for LRS-BI model by using 
the temporal component of field equation i.e.~equation (\ref{eom1}) along 
with equations (\ref{Q}) and (\ref{Qnew}). The expression  of Hubble 
parameter takes the form:
\begin{equation}\label{Hubble}
H = \sqrt{\frac{1}{6f_{Q}}\big[\frac{f(Q)}{2}-k\rho \big]+\frac{\sigma^2}{3}}.
\end{equation}
With the help of this Hubble parameter expression, we can also derive other 
cosmological parameters like deceleration parameters, luminosity distance, 
distance modulus in $f(Q)$ gravity for the model.

Now, the present scenario gives us a system of three equations (\ref{eom1},\ref{eom2},\ref{eom3}) with five unknowns $H_x,H_y,\rho, p_x,p_y$ and the model $f(Q)$. So even for model specific analysis, we need two additional assumptions to close the system. We take the first condition in form of the barotropic EOS, $p=\omega \rho$ and the second as the simplest physically relatable condition $\sigma^2\propto\theta^2$. This latter condition is well-known in anisotropy-based cosmology literature \cite{B1,B2,B3,B4,B5,B6,B7} and also very recently being utilised in probing $f(Q)$ theory in early universe \cite{jcap/AD}, in the same article it is also showed that in radiation era, a quadratic model $f(Q) = f_0 Q^2$ reproduces the above condition. Under this proportionality, we have the relation
\begin{equation}\label{H_directional}
    H_x=\lambda H_y,
\end{equation}
where $\lambda$ is a constant. Therefore, the average Hubble parameter becomes 
\begin{align}
H=\frac{(2+\lambda)}{3}H_y.
\end{align}
Accordingly, equations 
(\ref{Q}, \ref{eom1}, \ref{eom2} and \ref{eom3}) take the forms as follows:
\begin{eqnarray}
    Q&=&\frac{18(1+2\lambda)}{(2+\lambda)^2}H^2,\\[5pt]
    k\rho &=&\frac{f(Q)}{2}-\frac{18(1+2\lambda)}{(2+\lambda)^2}H^2f_Q,\label{em1}\\[5pt]
    kp_x&=&-\frac{f(Q)}{2}+\frac{6}{2+\lambda}f_Q \big[3H^2+\dot{H} \big]+\frac{216(1+2\lambda)}{(2+\lambda)^3}H^2\dot{H}f_{QQ},\label{em2}\\[5pt]
    kp_y&=&-\frac{f(Q)}{2}+3f_Q\left(\frac{1+\lambda}{2+\lambda}\right)\left[3H^2+\dot{H}\right]+\frac{108(1+\lambda)(1+2\lambda)}{(2+\lambda)^3}H^2\dot{H}f_{QQ}\label{em3}.
\end{eqnarray}
For equation (\ref{H_directional}), the condition derived for the pressure 
isotorpy i.e. equation (\ref{fh}) reduce to
 \begin{equation}
 \frac{k}{a^3} = \big(\frac{\lambda - 1}{\lambda + 2} \big)f_Q H.
 \end{equation}
Similarly, the effective EoS and the DE EoS given by equations (\ref{eoseff}) 
and \eqref{eosde} respectively can be written as
 \begin{align}\label{omega_eff_new}
 \omega^{eff} & = \frac{p+\frac{f(Q)}{2}-3H^2f_Q-(\frac{6}{2+\lambda}) H\dot{Q}f_{QQ}}{\rho-\frac{f(Q)}{2}+3H^2f_Q-\sigma^2 f_Q},\\[5pt] \label{eosde_new}
 \omega^{DE} & = -\frac{\frac{f(Q)}{2}-3H^2f_Q-(\frac{6}{2+\lambda}) H\dot{Q}f_{QQ}}{\frac{f(Q)}{2}-3H^2f_Q+\sigma^2 f_Q}.
 \end{align}
 Further, the value of $\dot{Q}$ can be written in the following form:
 \begin{equation}\label{Q_derivative}
 \dot{Q} = 36H(1+2\lambda)\Big[\frac{p(2+\lambda)+\frac{f(Q)(2+\lambda)}{2}-{18f_{Q}H^2}}{6f_{Q}(2+\lambda)^{2}+216(1+2\lambda)H^2f_{QQ}} \Big].
 \end{equation}
 The above expression of $\dot{Q}$ leads to the required form of effective 
EoS to be used for the graphical analysis purpose for various $f(Q)$ models.
 
For the relation between the directional Hubble parameters in equation 
(\ref{H_directional}), the relation between directional scale factors can be 
obtained as
\begin{equation}
A = c B^{\lambda}.
\end{equation}
Here, c is a constant and for simplification we have taken $c=1$. Thus, the 
average scale factor takes the form: $a = B^{\frac{\lambda+2}{3}}$ and hence 
the total energy density of the Universe can be found as 
 \begin{equation}
 \rho = \rho_{0} B^{-(1+\omega)(2+\lambda)},
 \end{equation}
{which follows from the energy-momentum conservation in the studied setting as shown in \cite{deepjc}}. However, the scale factor parameter is not 
an observational parameter. Therefore it is better to express all the 
important cosmological parameters in terms of cosmological redshift. As the 
direct observational data of various parameters like Hubble parameter, 
distance modulus etc.~against cosmological redshift are available, so by 
evaluating the theoretical expressions of all the cosmological parameters in 
terms of cosmological redshift, one can compare theoretical results with the
observational data. So, if $z$ be the  cosmological redshift along the 
$y$-direction in the LRS-BI Universe, then the average scale factor $a$ can 
be derived from the metric (\ref{metric}) in terms of $z$ as
\begin{equation}
a = (AB^2)^{\frac{1}{3}} = (1+z)^{-\frac{(2+\lambda)}{3}}, 
\end{equation}
where $B = (1+z)^{-1}$ and $A = (1+z)^{-\lambda}$.
All the cosmological parameters in the following sections will be expressed in terms of  cosmological redshift $z$.
 
\section{LRS-BI Universe with two different $f(Q)$ gravity models}\label{sec5}
In this section, we try to obtain various cosmological parameters for two 
different form of $f(Q)$-gravity models to develop basic understanding about
the LRS-BI Universe and also try to find out the role of anisotropy in the 
early stage of the Universe, or in the near past as follows:

\subsection{$f(Q) = -(Q+2\Lambda)$ model}
First we take a simple form of $f(Q)$, i.e.~$f(Q) = -(Q+2\Lambda)$, 
where $\Lambda$ is the cosmological constant.
{This model is just an extension of the $f(Q)=-Q$ model, which mimics 
the classical GR results in $f(Q)$-gravity \cite{deepjc,Q6}. Here we have added 
the additional cosmological constant part to get more physical and realistic 
results in conformity with the present day cosmology. It is to be noted here 
that this model in isotropic cosmology is nothing but the $\Lambda$CDM model
  \cite{Q6,gaurav_2022}. Here we are interested to see its role in LRS-BI 
Universe.}
  Now, the expressions of 
effective EoS, DE EoS and Hubble parameter given by equations 
(\ref{omega_eff_new}), (\ref{eosde_new}) and (\ref{Hubble}) respectively 
take the forms for this model as
\begin{align}
\omega^{eff} & = \frac{\frac{1}{3}\Omega_{r0} (1+z)^{\frac{4(2+\lambda)}{3}}-\Omega_{\Lambda 0} + \Omega_{\sigma 0} (1+z)^{2(2+\lambda)}}{\Omega_{mo} (1+z)^{(2+\lambda)}+\Omega_{r0} (1+z)^{\frac{4(2+\lambda)}{3}}+\Omega_{\Lambda 0}},\\[5pt]
\label{eosde_new1} 
\omega^{DE} & = - 1 + \frac{\Omega_{\sigma 0}}{\Omega_{\Lambda 0}} (1+z)^{2(2+\lambda)},\\[5pt]
 H & = H_{0} \sqrt{\Omega_{mo} (1+z)^{(2+\lambda)}+\Omega_{r0} (1+z)^{\frac{4(2+\lambda)}{3}}+\Omega_{\Lambda 0}+\Omega_{\sigma 0} (1+z)^{2(2+\lambda)}}.
\end{align}
Here $H_0$ is the current Hubble parameter, $\Omega_{m0} = 
8\pi \rho_{m0}/3H_{0}^{2}$ is the density parameter for the matter 
content, $\Omega_{r0} = 8\pi  \rho_{r0}/3H_{0}^{2}$ is the density parameter 
for the radiation content, $\Omega_{\Lambda 0} = \Lambda/3H_{0}^{2}$ is the 
density parameter for the vacuum energy and $\Omega_{\sigma 0} = 
\sigma_{0}^{2}/3H_{0}^{2}$ is the density parameter for the anisotropy of the 
present Universe \cite{sarmah}. The term $\rho_{m0}$ and $\rho_{r0}$ are 
current values of matter density and radiation density of the Universe 
respectively. Again, the term $\sigma_{0}^{2}$ in $\Omega_{\sigma 0}$ is the 
current value of shear scalar which is related to  the shear scalar 
$\sigma^2$ by the relation $\sigma^{2}= \sigma_0^2 a^{-6}$  and 
$\Omega_{\sigma 0} \leq 10^{-15}$ \cite{akarsu_2019,sarmah}.

\begin{figure}[!h]
\centerline{
\includegraphics[scale = 0.35]{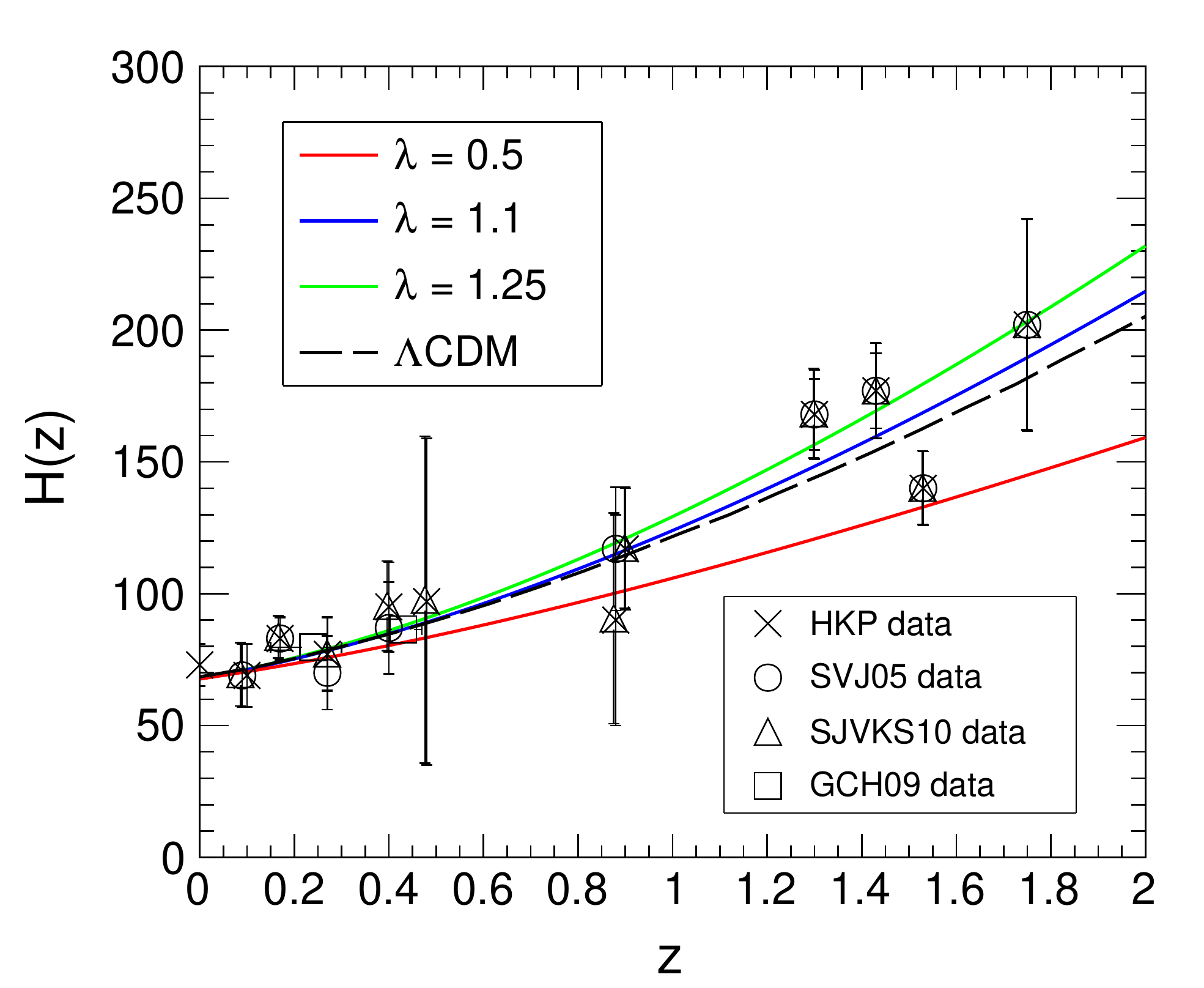}}
\vspace{-0.2cm}
\caption{Behaviour of Hubble parameter $H(z)$ with respect to $z$ for
different values of the parameter $\lambda$ as predicted by the model
$f(Q) = -(Q+2\Lambda)$. The model predictions are in comparison with the
four sets of Hubble parameter data, viz, HKP and SVJO5 data \cite{simon_2005}, 
SJVKS10 data \cite{stern_2010} and GCH09 data \cite{hui_2009}.}
\label{fig1a}
\end{figure}

Similarly, the deceleration parameter ($q$) and luminosity  distance ($d_L$) 
for the given $f(Q)$ model can be written as
\begin{equation}\label{q_1}
q = 1 - \frac{\lambda}{2} + (\frac{2+\lambda}{2})\big[ \frac{\Omega_{r0} (1+z)^{\frac{4(2+\lambda)}{3}}-\Omega_{\Lambda 0} + \Omega_{\sigma 0} (1+z)^{2(2+\lambda)}}{\Omega_{mo} (1+z)^{(2+\lambda)}+\Omega_{r0} (1+z)^{\frac{4(2+\lambda)}{3}}+\Omega_{\Lambda 0}+\Omega_{\sigma 0} (1+z)^{2(2+\lambda)}}\big], 
\end{equation}
\begin{equation}\label{d_L_1}
 d_{L} = \frac{(\lambda+2)(1+z)^{\frac{2+\lambda}{3}}}{3H_{0}}\int_{0}^{\infty} \big[\frac{(1+z)^{\frac{\lambda-1}{3}}}{\sqrt{\Omega_{mo} (1+z)^{(2+\lambda)}+\Omega_{r0} (1+z)^{\frac{4(2+\lambda)}{3}}+\Omega_{\Lambda 0}+\Omega_{\sigma 0} (1+z)^{2(2+\lambda)}}}\big]dz. 
 \end{equation}
With the help of equation (\ref{d_L_1}) we can calculate the values of the plot reduce to
distance modulus for different values of $z$ by using the relation,
 \begin{equation}\label{Dm}
 D_m = 5 \log{d_L} + 25.     
 \end{equation}
To see and understand the behaviours of all these cosmological parameters at 
different stages of evolution of the Universe, we have to plot these 
parameters against the cosmological redshift $z$. For this purpose we have to 
choose a reliable range of values of the model parameter i.e.~$\lambda$. To 
estimate this range of $\lambda$ we have plotted the Hubble parameter for 
different values of $\lambda$ along with four sets of observational data, 
viz.~HKP and SVJO5 data \cite{simon_2005}, SJVKS10 data \cite{stern_2010} 
and GCH09 data \cite{hui_2009} on the Hubble parameter and also  with the 
$\Lambda$CDM model prediction  up to $z = 2$ as shown in Fig.~\ref{fig1a}. The 
plot shows consistent results with the observational data for values of 
$\lambda$ within the range of 0.5 to 1.25 with $\lambda \ne 1$. 
It is to be noted that we have used Planck 2018 results \cite{Planck_2018} on the 
cosmological parameters for all numerical calculations.

\begin{figure}[!h]
\centerline{
\includegraphics[scale = 0.35]{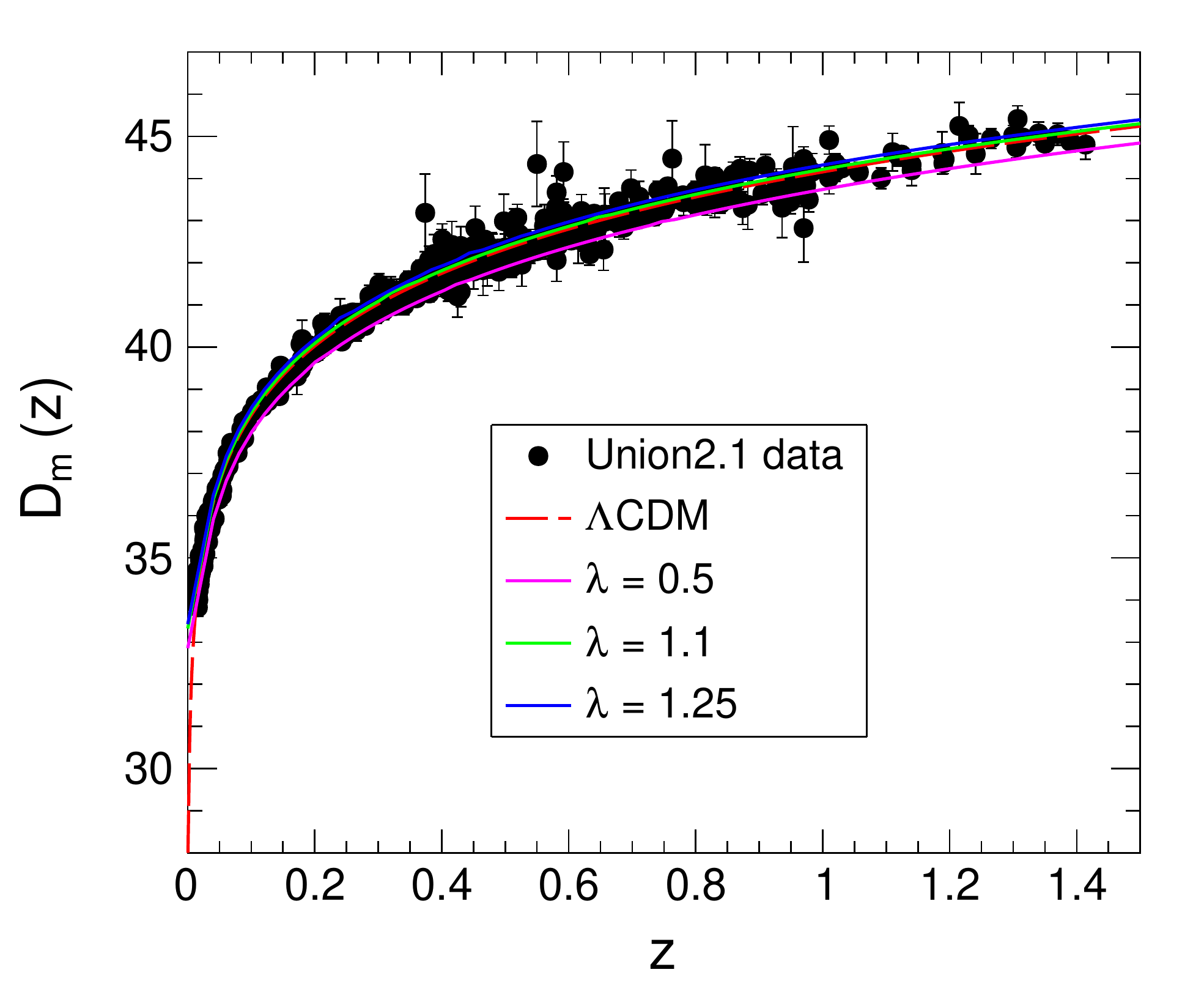}}
\vspace{-0.2cm} 
\caption{Behaviour of distance Modulus $D_{m}$ with respect to cosmological 
redshift $z$ for different values of the parameter $\lambda$ as predicted by 
the model $f(Q) = -(Q+2\Lambda)$. The model predictions are in comparison with 
the SCP Union 2.1 observational data \cite{suzuki_2012}.}
\label{fig1b}
\end{figure}

\begin{figure*}[!h]
\centerline{
  \includegraphics[scale = 0.28]{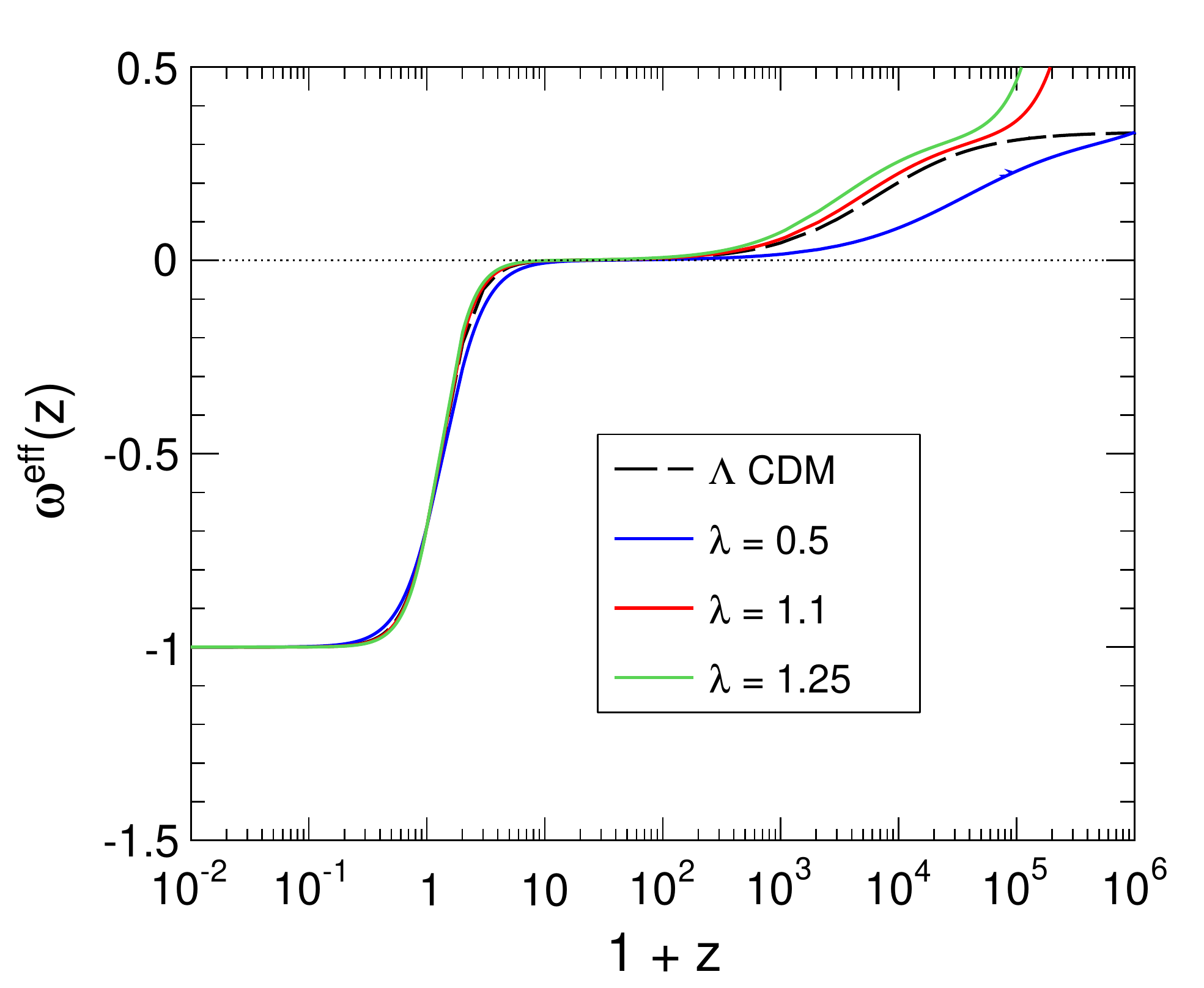}\hspace{0.5cm}
  \includegraphics[scale = 0.28]{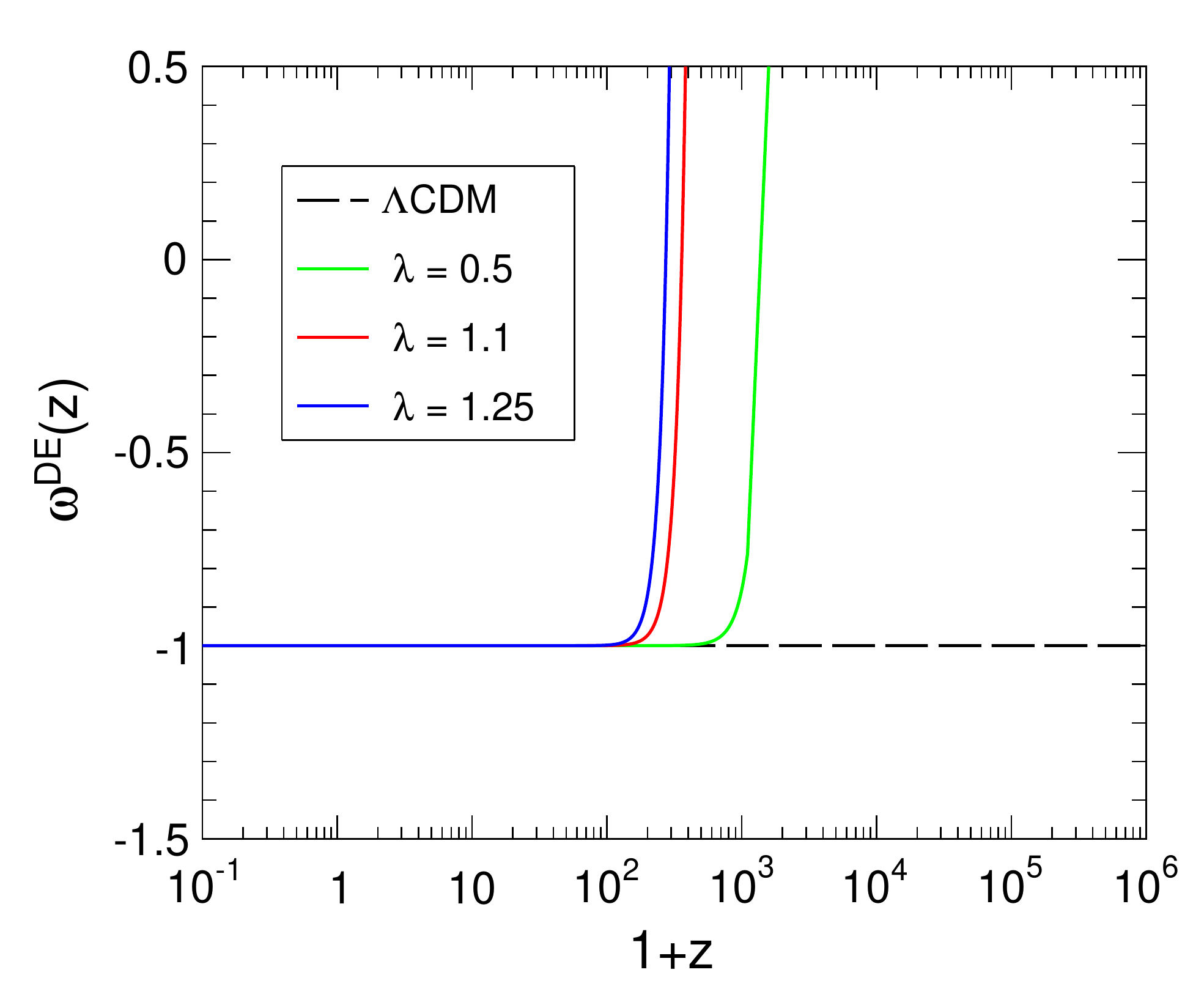}\hspace{0.5cm}  
  \includegraphics[scale = 0.28]{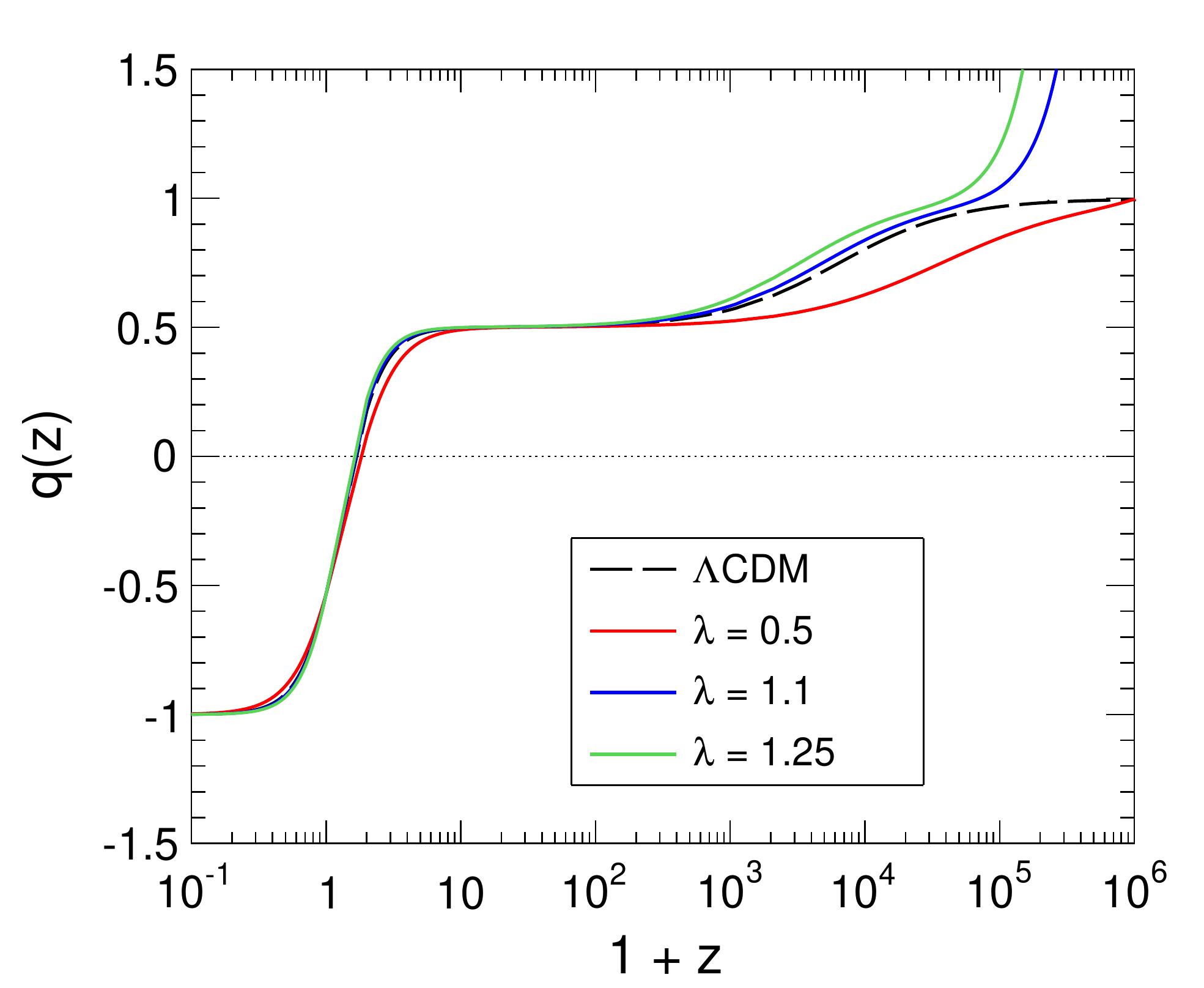}}
  \vspace{-0.2cm} 
\caption{Behaviour of effective EoS parameter $\omega^{eff}$ (left), DE EoS 
parameter $\omega^{DE}$ (middle) and deceleration parameter $q$ (right) 
against the cosmological redshift $z$ for different values of parameter 
$\lambda$ as predicted by the model $f(Q) = -(Q+2\Lambda)$ in comparison with 
that of $\Lambda$CDM model.}
\label{fig1c}
\end{figure*}

Similarly, have plotted distance modulus $D_{m}$ against cosmological redshift 
$z$ in Fig.~\ref{fig1b} for the same set of $\lambda$ values as considered 
in the Hubble parameter plot, along with the SCP Union 2.1 observational 
data \cite{suzuki_2012} and the corresponding $\Lambda$CDM model prediction. 
This plot also shows the consistent agreement with the observational data.
Based on the above two plots we have constrained the values of the model 
parameter $\lambda$ within the range 0.5 to 1.25. And hence using the above 
considered set of values of $\lambda$, we have plotted the effective EoS 
parameter $\omega^{eff}$, DE EoS parameter $\omega^{DE}$ and deceleration 
parameter $q$ against the cosmological redshift $z$ in Fig.~\ref{fig1c}.
We have seen that both $\omega^{eff}$ and $q$ deviate significantly from the 
$\Lambda$CDM case at higher values of $z$ for all the three anisotropic cases. 
However the deviation decreases to some level as the $\lambda$ tends to unity
as expected. In the case of $\omega^{DE}$ the deviation from the standard 
cosmology is more vivid, prominently abrupt and takes place at smaller values 
of $z$ (depending on the value of $\lambda$) in comparison to the cases of 
$\omega^{eff}$ and $q$. In this case the model agrees with $\Lambda$CDM upto
$z\sim 10^2$, but for $z>10^2$ the model shows sharp increase of the value of
$\omega^{DE}$ towards the positive side. This sharp and abrupt behaviour
of $\omega^{DE}$ is due to the contribution of the anisotropic factor in
the early Universe as clear from equation \eqref{eosde_new1}. For higher $z$ 
values this factor becomes substantial than the $\Lambda$CDM part. Thus this 
model shows that the anisotropy was prominent at the early stage of the 
Universe as the $\lambda$ deviated from one. 
 
\subsection{$f(Q) = - \alpha Q^{n}$ model}
This is the second model we consider in this study, which is a power law model 
with a constant multiplier $\alpha$ and with the exponent $n$. 
{The major advantage of the power law model is that it is mathematically
 simple yet powerful \cite{Hammad_2014}. Also the power law model can explain 
the late time cosmic acceleration and is also consistent with BBN constrains
 \cite{Fotios_2022}. The study of isotropic cosmology in $f(Q)$ theory while
considering power law model is found in 
Refs.~\cite{Khyllep_2022,Q6,capozziello_2022}.}
 For this model 
we can express $f(Q)$ and its derivatives in terms of cosmological density 
parameters as follows 
\begin{equation}\label{fq}
f(Q) = -\alpha Q^{n} = -\alpha\left[\frac{3^{\frac{1}{n}}H_{0}^{\frac{2}{n}}[\Omega_{mo} (1+z)^{(2+\lambda)}+\Omega_{r0} (1+z)^{\frac{4(2+\lambda)}{3}}+\Omega_{\Lambda 0}]}{(n-\frac{1}{2})^{\frac{1}{n}}\alpha^{\frac{1}{n}}}\right]^{n},
\end{equation}
\begin{equation}
 f_{Q} = -\alpha n\left[\frac{3^{\frac{1}{n}}H_{0}^{\frac{2}{n}}[\Omega_{mo} (1+z)^{(2+\lambda)}+\Omega_{r0} (1+z)^{\frac{4(2+\lambda)}{3}}+\Omega_{\Lambda 0}]}{(n-\frac{1}{2})^{\frac{1}{n}}\alpha^{\frac{1}{n}}}\right]^{n-1},
\end{equation}
\begin{equation}
 f_{QQ} = -\alpha n(n-1)\left[\frac{3^{\frac{1}{n}}H_{0}^{\frac{2}{n}}[\Omega_{mo} (1+z)^{(2+\lambda)}+\Omega_{r0} (1+z)^{\frac{4(2+\lambda)}{3}}+\Omega_{\Lambda 0}]}{(n-\frac{1}{2})^{\frac{1}{n}}\alpha^{\frac{1}{n}}}\right]^{n-2}.
 \end{equation}
In above equations we have used the expression of Hubble parameter for this 
model, which takes the form:
\begin{equation}
H = \sqrt{\frac{3^{\frac{1}{n}}H_{0}^{\frac{2}{n}}[\Omega_{mo} (1+z)^{(2+\lambda)}+\Omega_{r0} (1+z)^{\frac{4(2+\lambda)}{3}}+\Omega_{\Lambda 0}]}{6(n-\frac{1}{2})^{\frac{1}{n}}\alpha^{\frac{1}{n}}}+H_{0}^{2}\Omega_{\sigma 0}(1+z)^{2(2+\lambda)}}.
\end{equation}
Now, we are in a position to derive and calculate the effective EoS parameter 
($\omega^{eff}$) and DE EoS parameter ($\omega^{DE}$) from equations 
(\ref{eoseff}) and ({\ref{eosde}}) respectively for this model by using above 
expressions for $f(Q),f_{Q}$ and $f_{QQ}$. Moreover, we can derive the 
deceleration parameter $q$ and luminosity distance $d_L$ for 
this model by using the following expressions:
 \begin{equation}
 q = -\big(1+\frac{\dot{H}}{H^2}\big),
 \end{equation}
 \begin{equation}\label{dL_powerlaw}
 d_{L} = \frac{(\lambda+2)(1+z)^{\frac{2+\lambda}{3}}}{3}\int_{0}^{\infty} \big[\frac{(1+z)^{\frac{\lambda-1}{3}}}{H(z)}\big]dz. 
 \end{equation}
Finally, using equation (\ref{dL_powerlaw}) in equation (\ref{Dm}), we can 
derive the expression of the distance modulus for this power law model.

\begin{figure}[!h]
\centerline{
  \includegraphics[scale = 0.28]{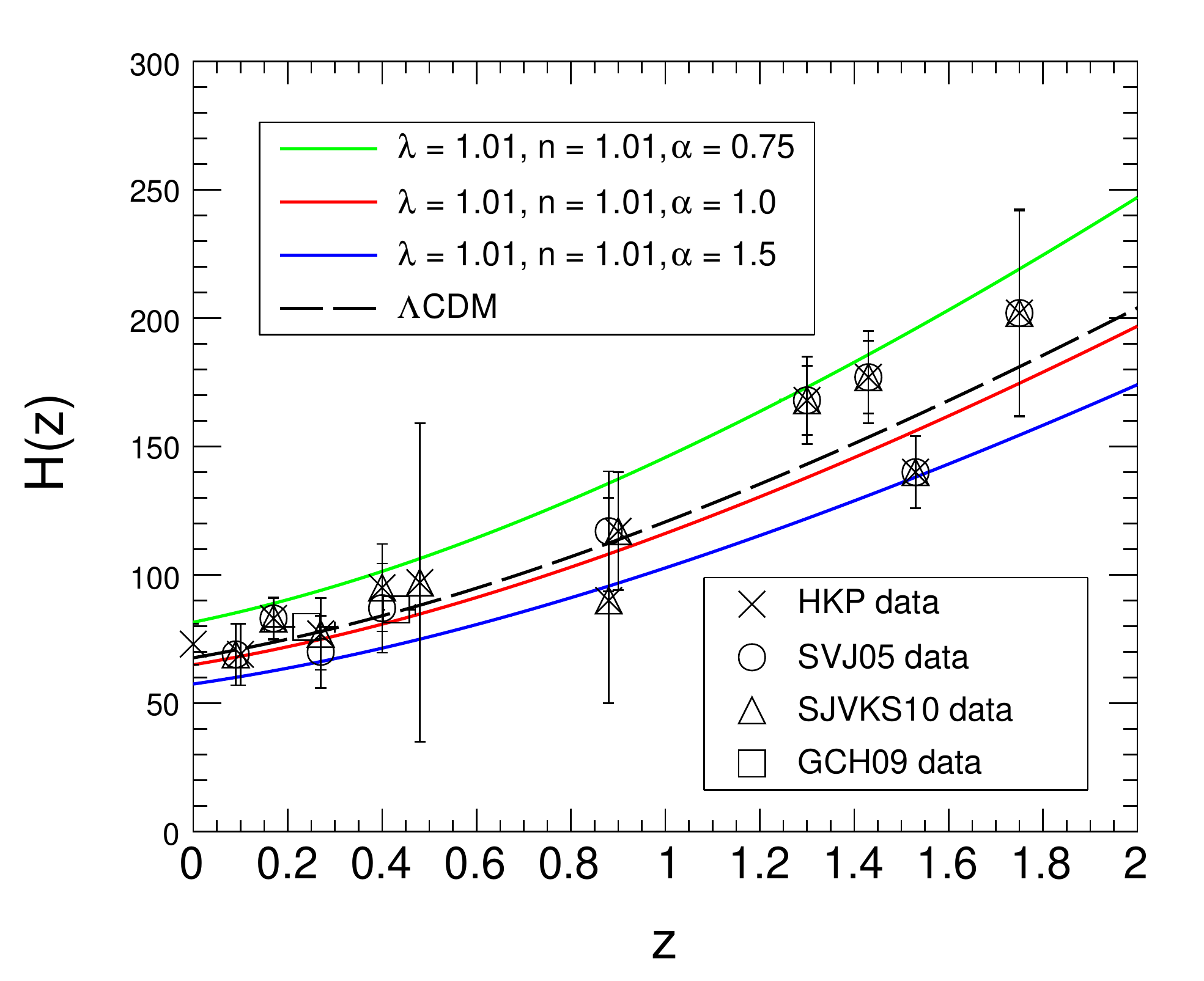}\hspace{0.25cm}
  \includegraphics[scale = 0.28]{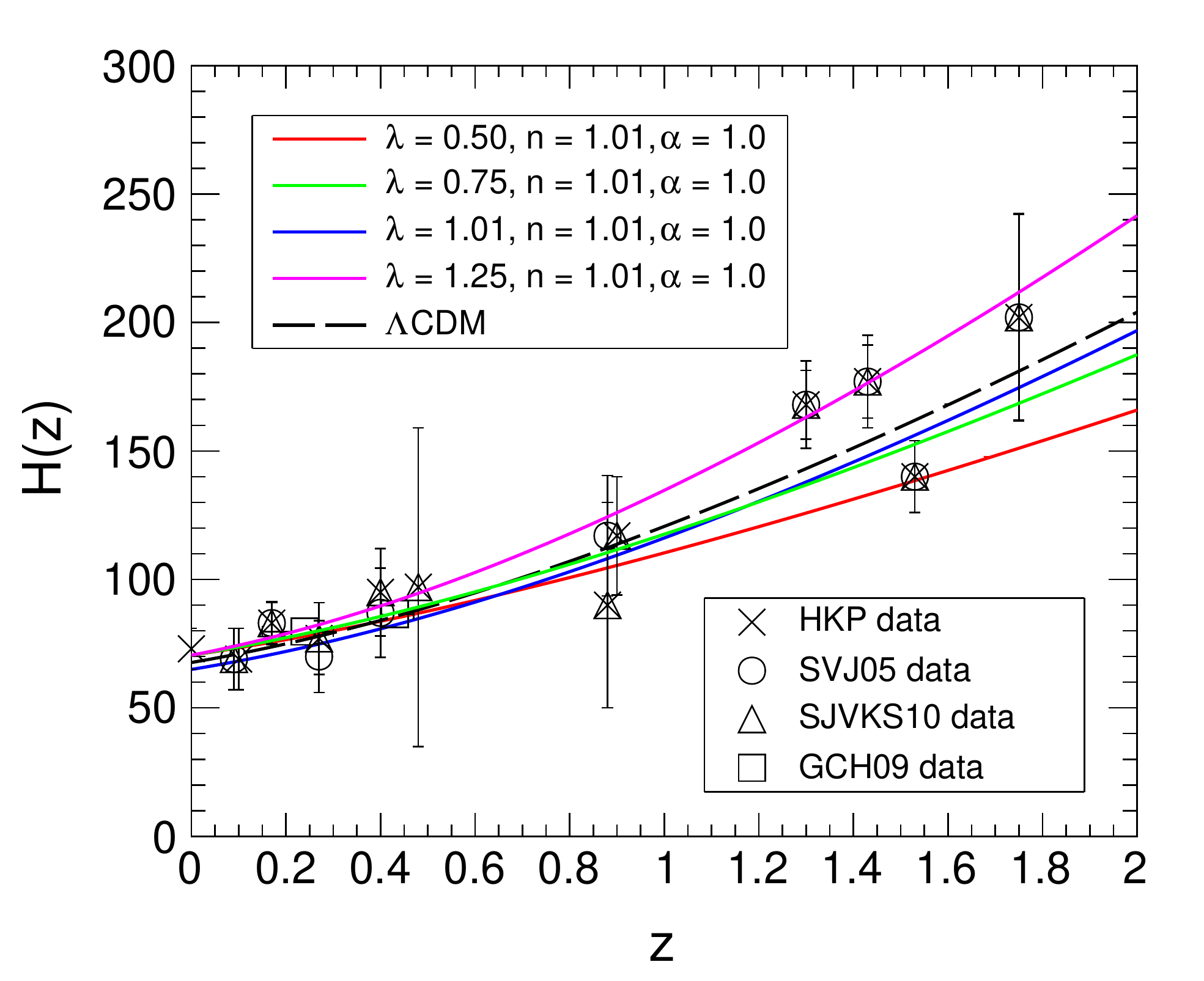}\hspace{0.25cm}
  \includegraphics[scale = 0.28]{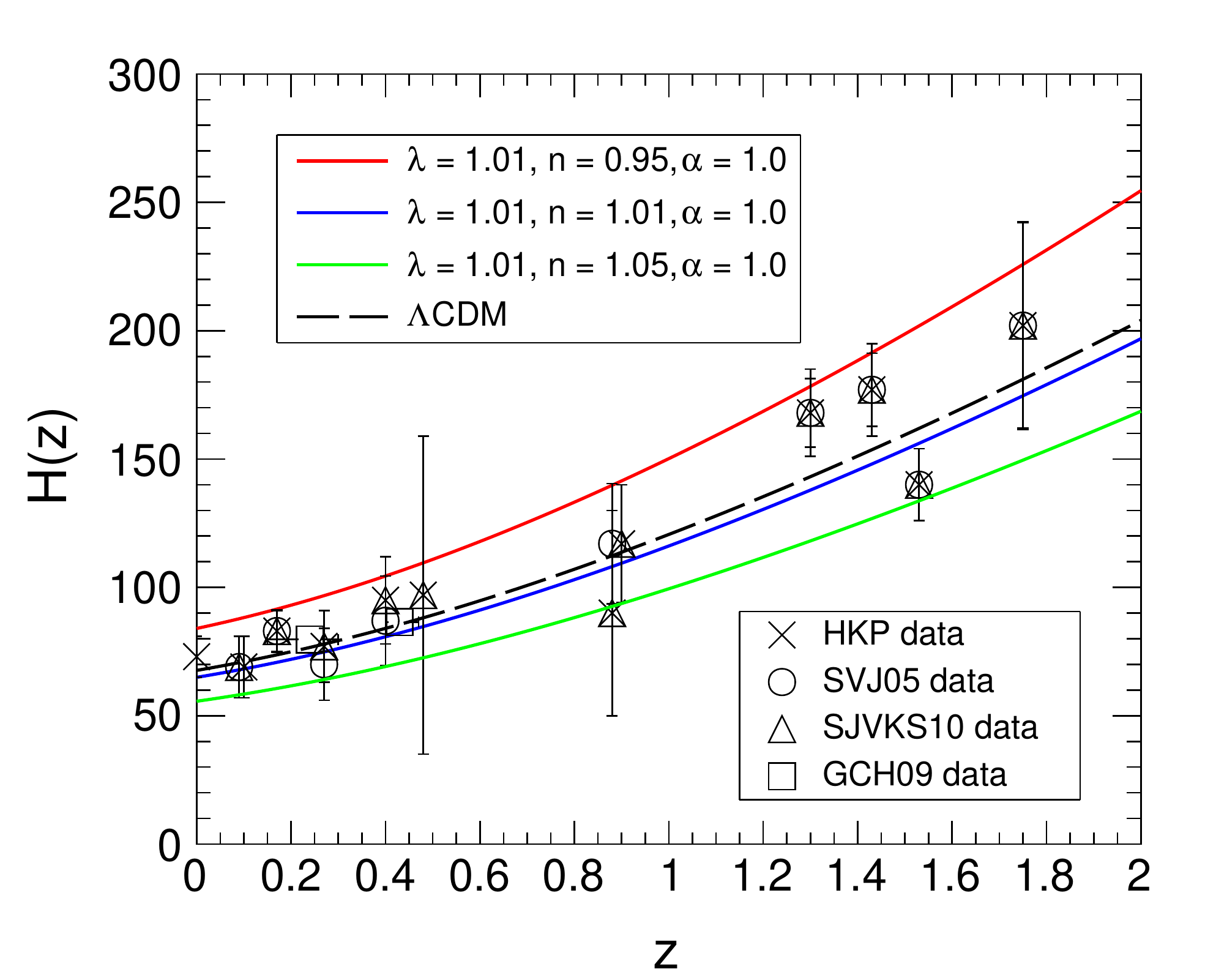}}
\vspace{-0.2cm}
\caption{Hubble parameter $H(z)$ against cosmological redshift $z$ for
different sets of values of model parameters $\lambda$, $\alpha$ and 
$n$ as predicted by the power law model. The model predictions are in 
comparison with the four sets of Hubble parameter data, viz, HKP and SVJO5 
data \cite{simon_2005}, SJVKS10 data \cite{stern_2010} and GCH09 
data \cite{hui_2009}.}
\label{fig2a}
\end{figure}

\begin{figure}[!h]
\centerline{
  \includegraphics[scale = 0.28]{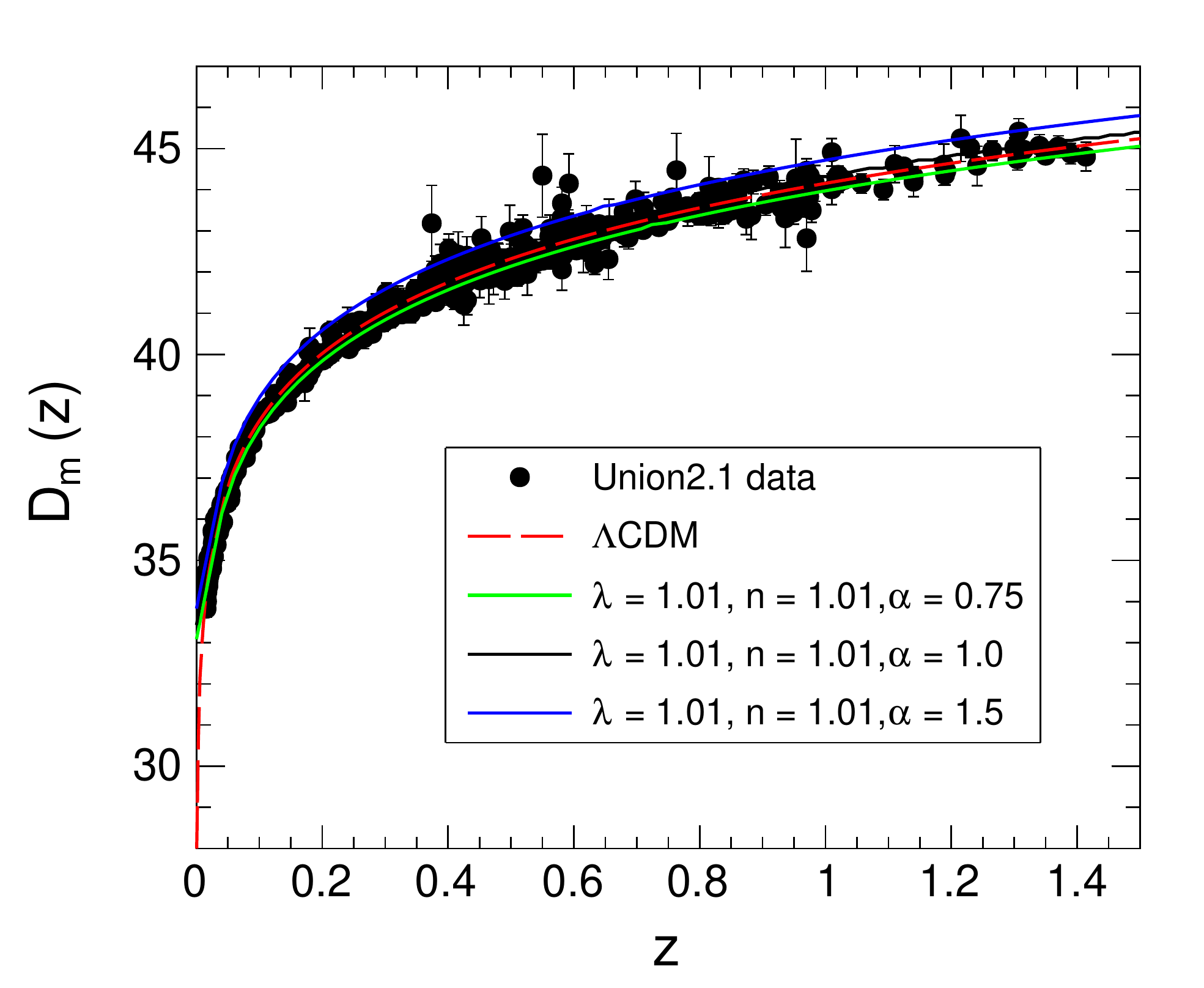}\hspace{0.25cm}
  \includegraphics[scale = 0.28]{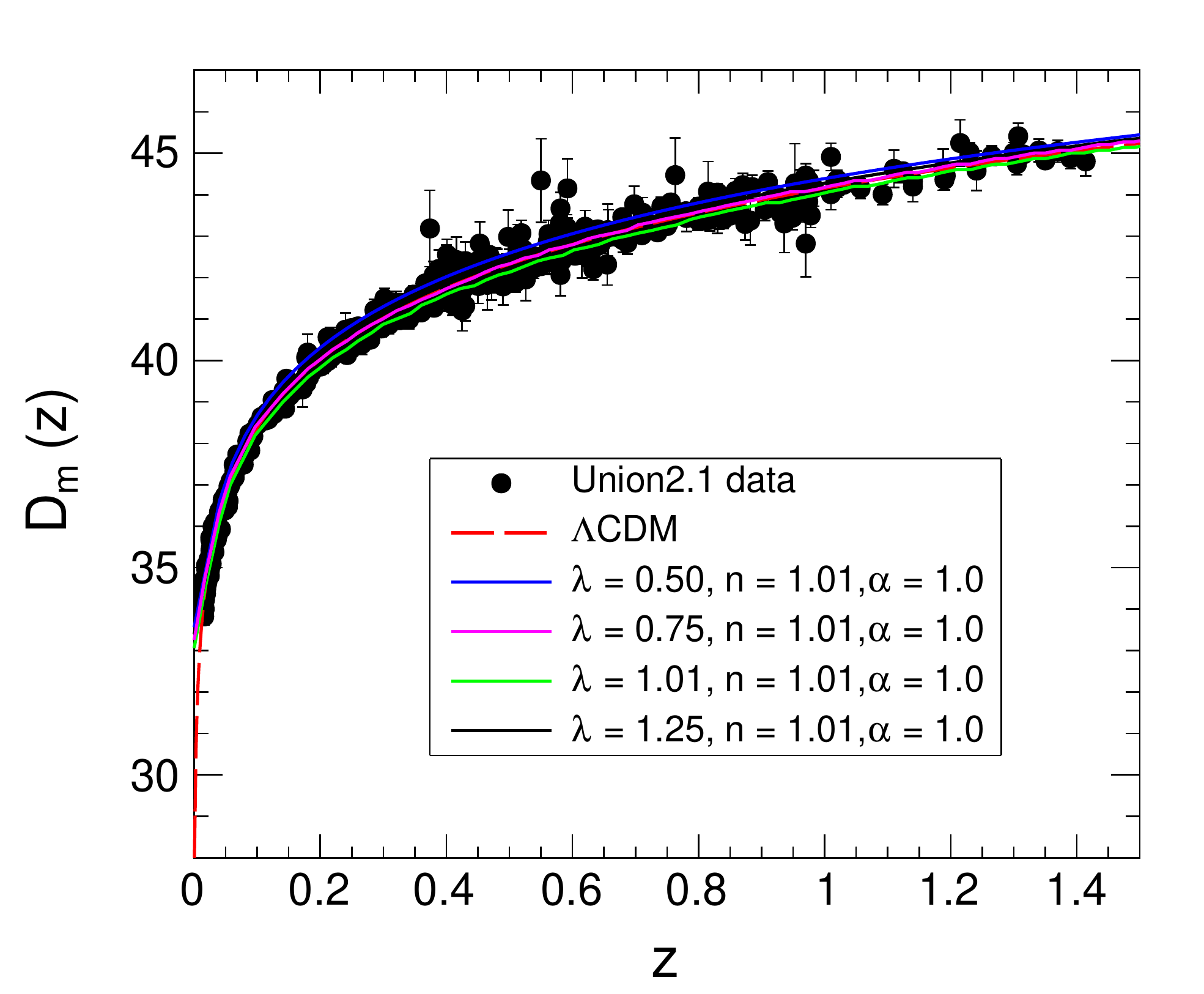}\hspace{0.25cm}
  \includegraphics[scale = 0.28]{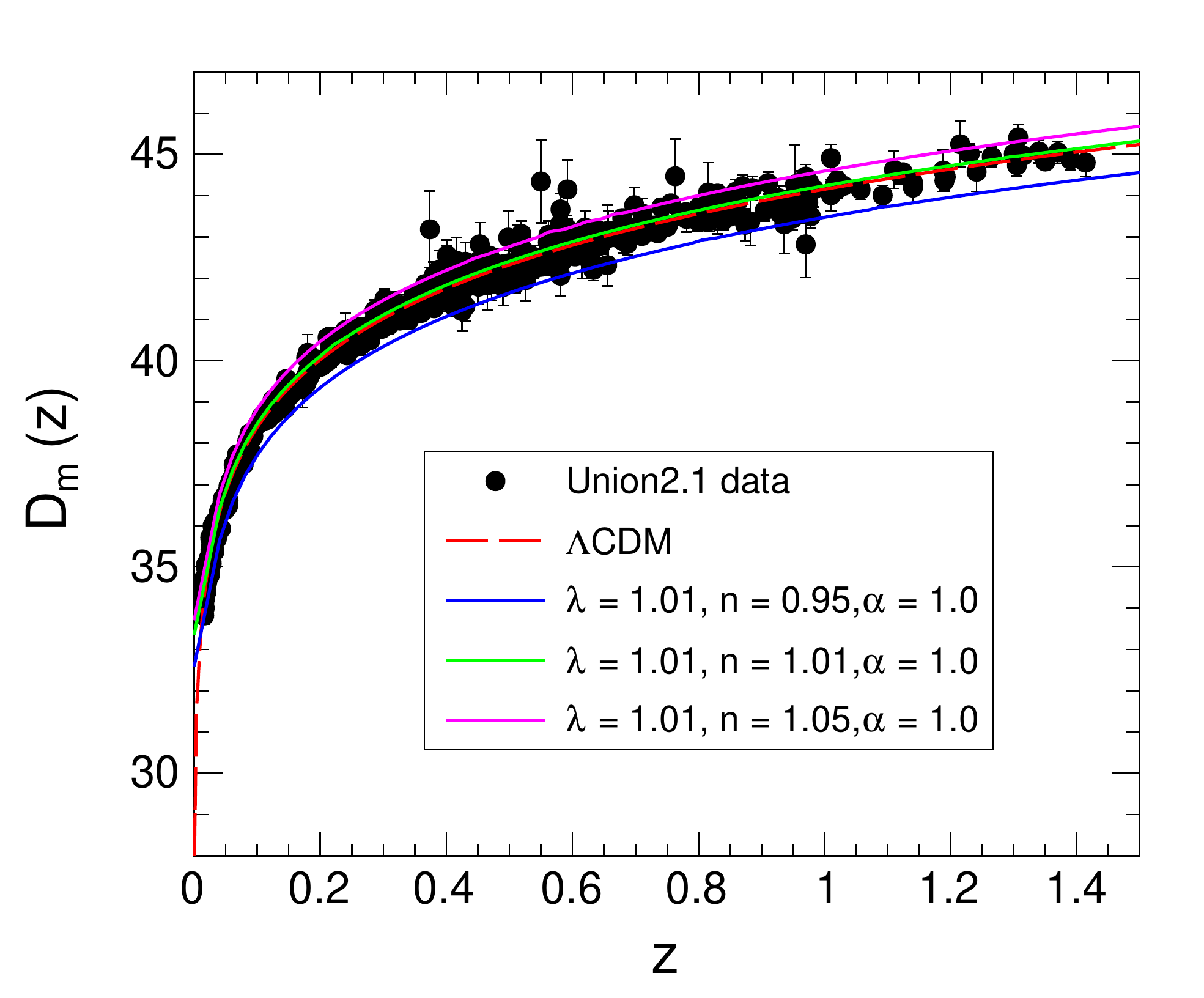}}
\vspace{-0.2cm}
\caption{ Distance modulus $D_{m}(z)$ against cosmological redshift $z$ 
for different sets of values of model parameters $\lambda$, $\alpha$ and 
$n$ as predicted by the power law model. The model predictions are in 
comparison with the SCP Union 2.1 observational data \cite{suzuki_2012}.}
\label{fig2b}
\end{figure}

It is clear that all the cosmological parameters for this power law model 
depend on three model parameters, viz.~$\lambda$, $\alpha$ and 
$n$. Similar to the case of previous model, to constrain these model 
parameters we have plotted the Hubble parameter and the distance modulus 
against cosmological redshift $z$ for different sets of values of these 
parameters along with the HKP, SVJO5, SJVKS10 and GCH09 observational data 
and the $\Lambda$CDM plot in Figs.~\ref{fig2a} and \ref{fig2b} respectively. 
From these plots we constrained the parameters $\alpha$ and $n$ respectively
as $0.75 \le \alpha \le 1.5$ and $0.95 \le n \le 1.05$ with $n \ne 1$. 
(2019)

\begin{figure}[!h]
\centerline{
  \includegraphics[scale = 0.28]{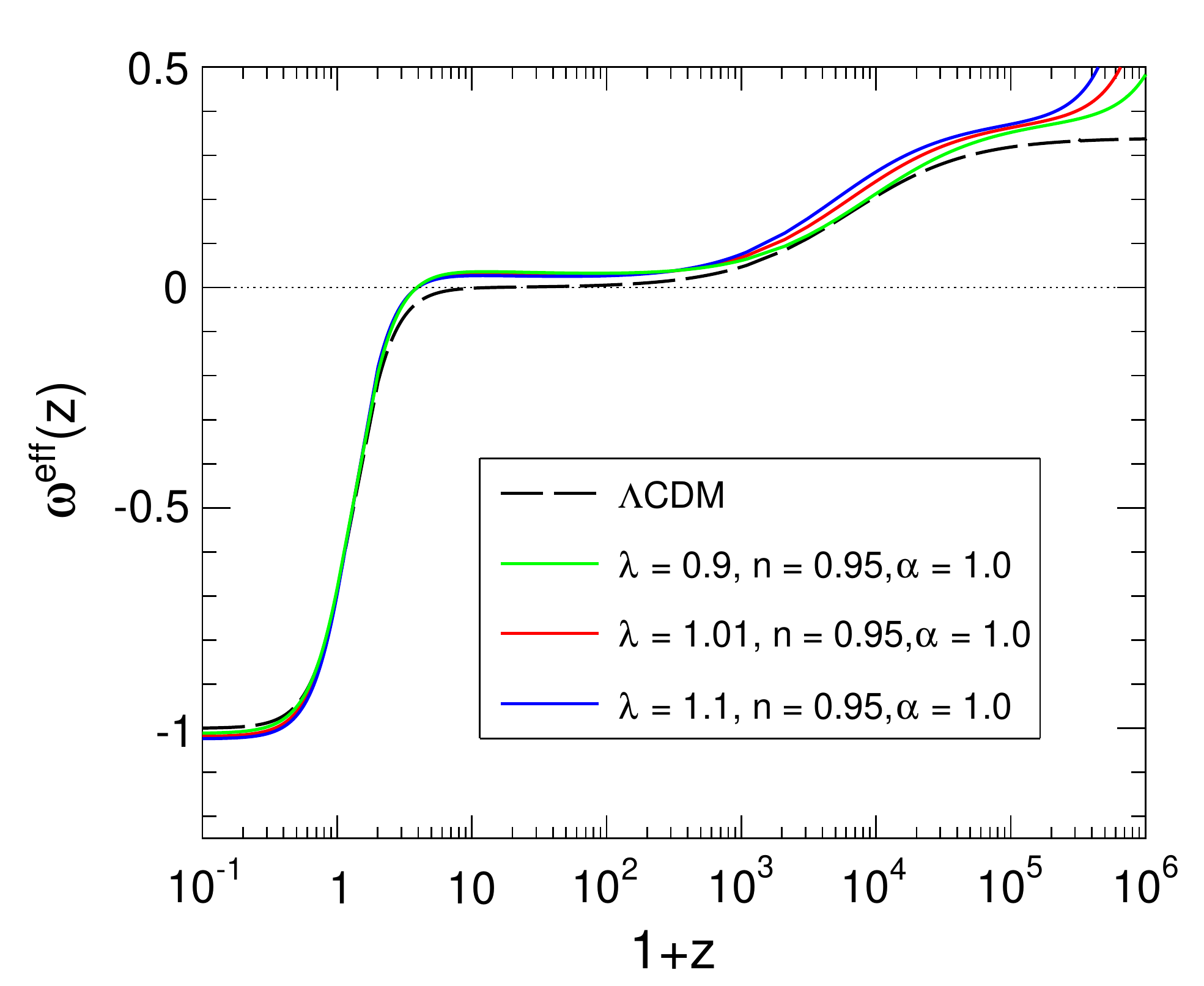}\hspace{0.25cm}
  \includegraphics[scale = 0.28]{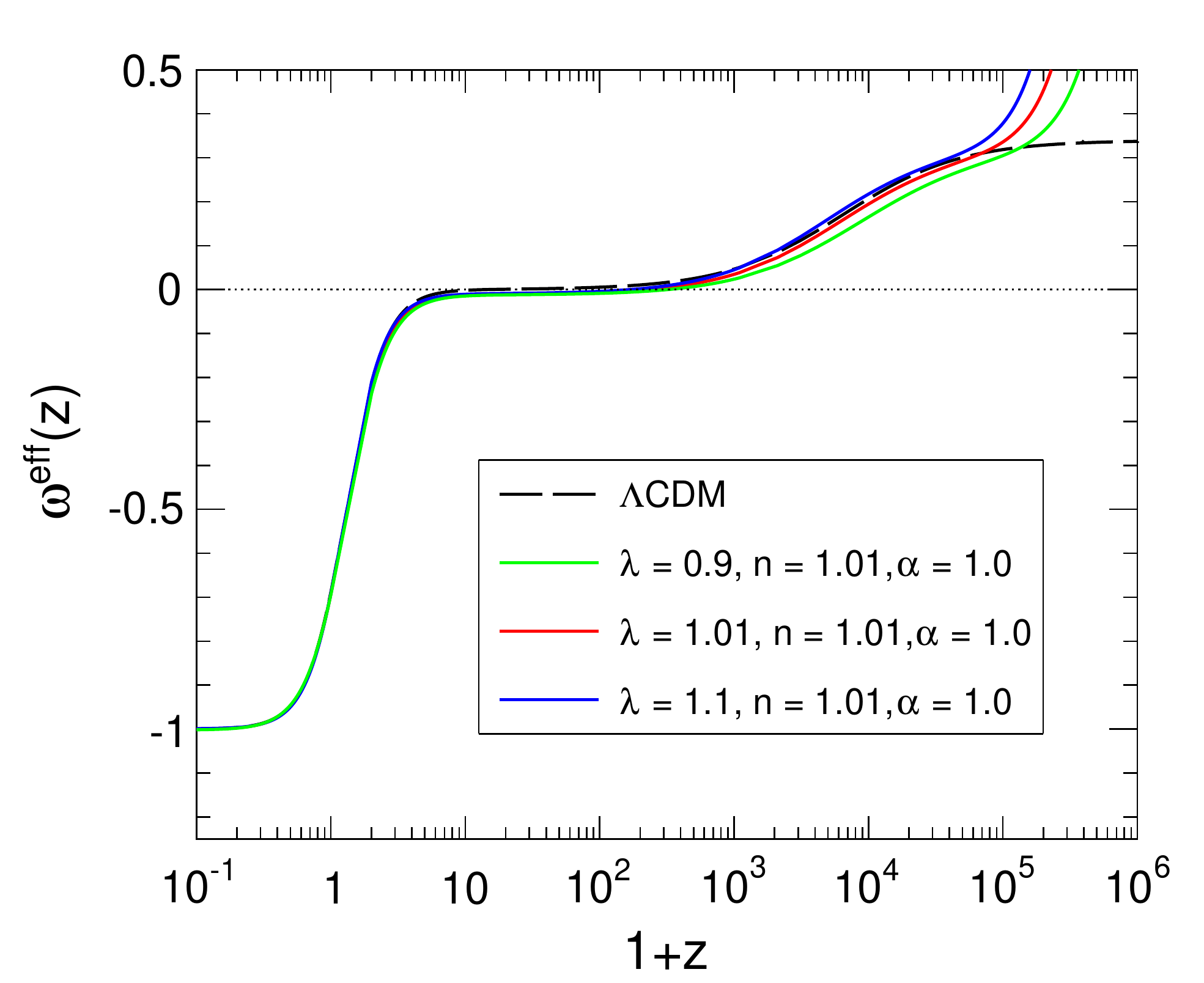}\hspace{0.25cm}
  \includegraphics[scale = 0.28]{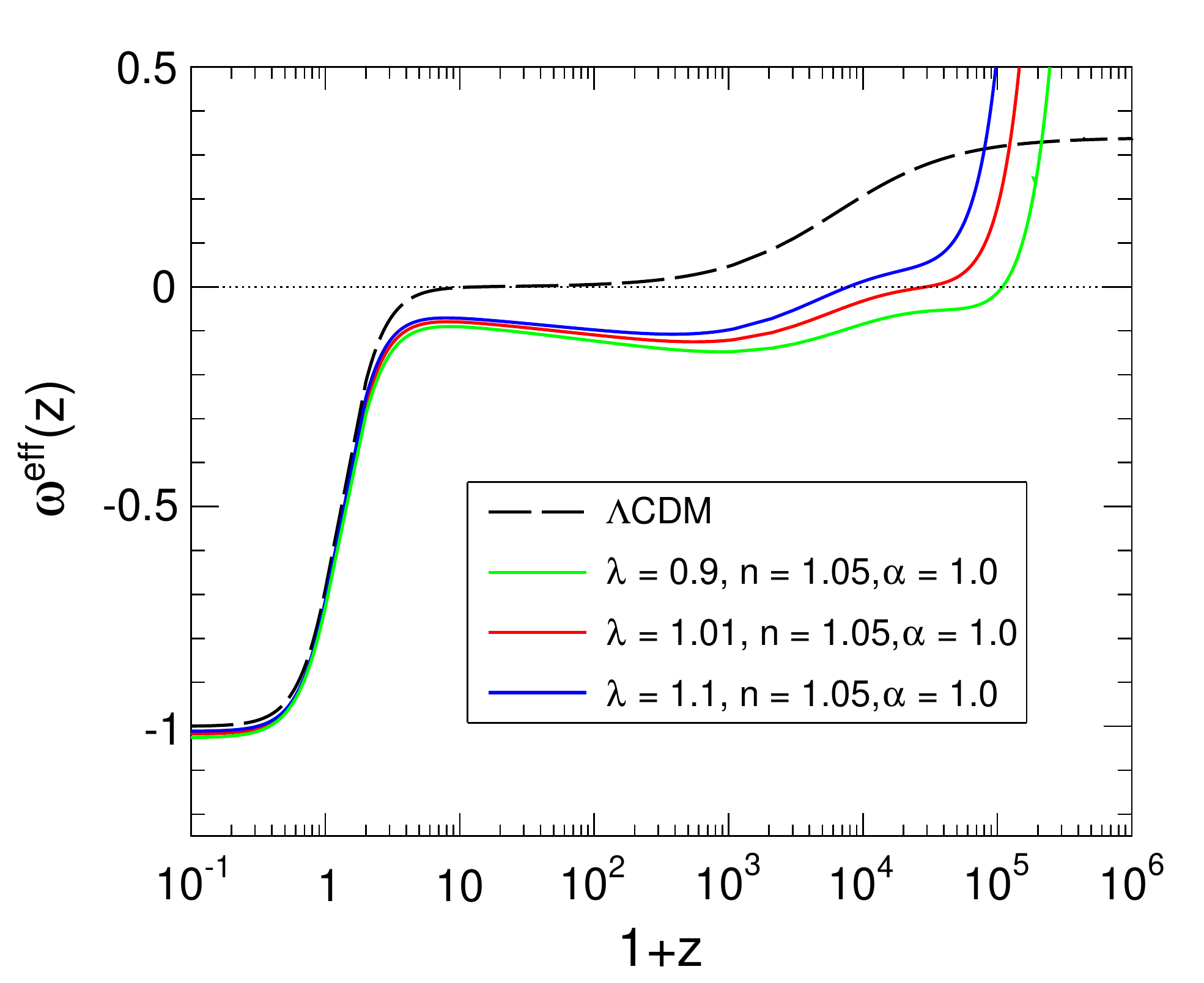}}
\vspace{-0.2cm}
\caption{Effective EoS parameter $\omega^{eff}$ against cosmological redshift 
$z$ for different sets of values of model parameters as predicted by the power
law model in comparison with that of $\Lambda$CDM model.}
\label{fig2c}
\end{figure}

\begin{figure}[!h]
\centerline{
  \includegraphics[scale = 0.28]{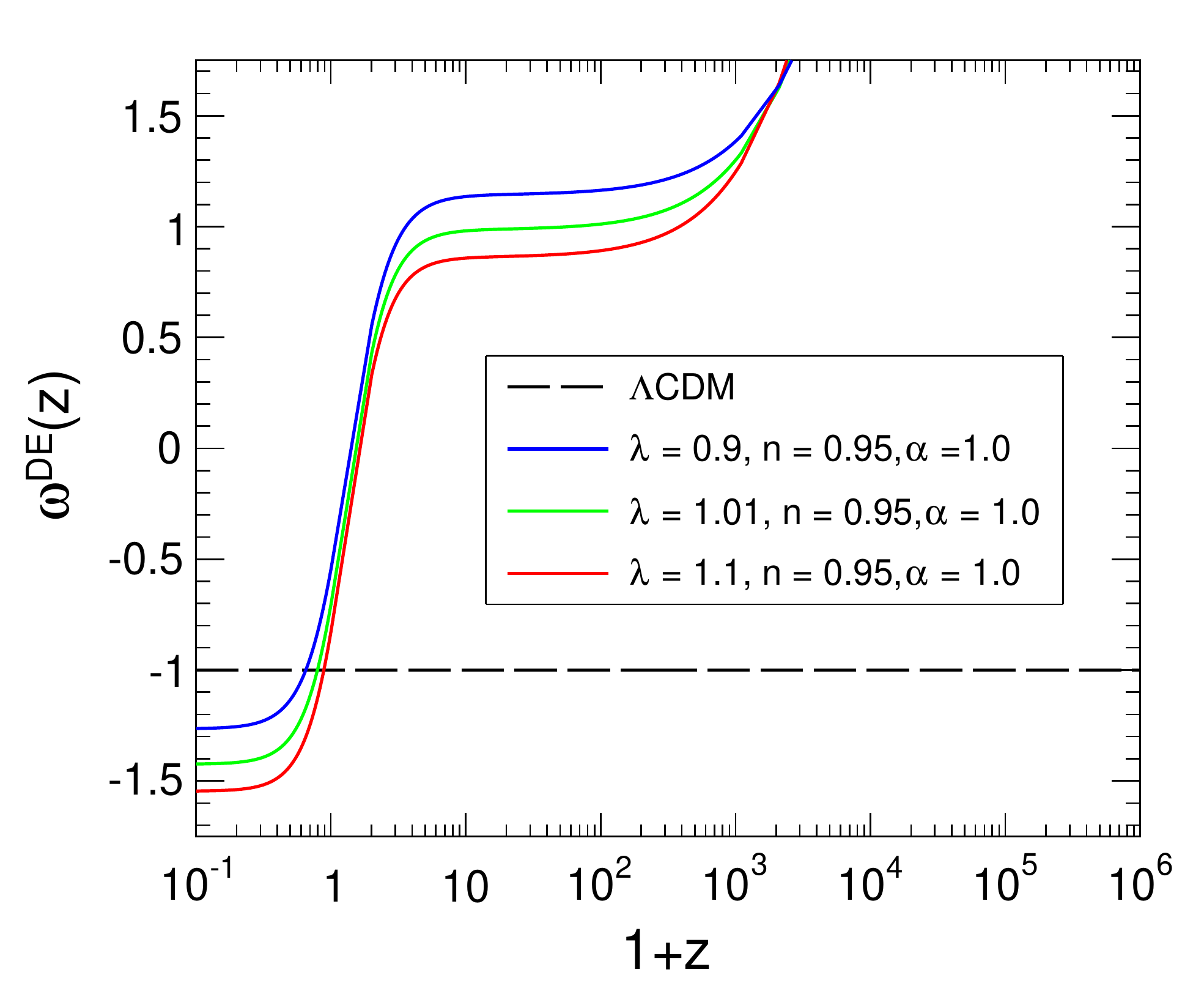}\hspace{0.25cm}
  \includegraphics[scale = 0.28]{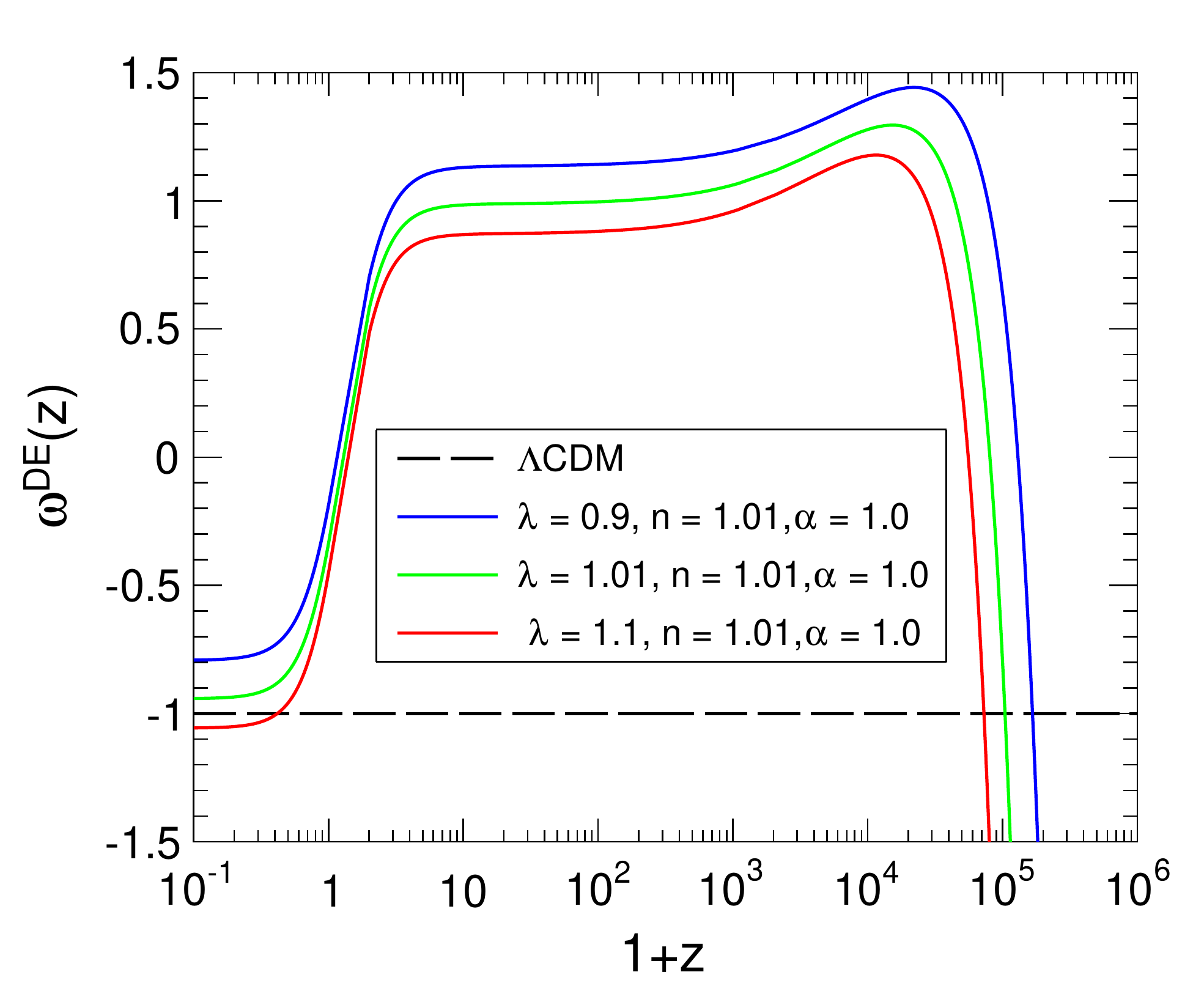}\hspace{0.25cm}
  \includegraphics[scale = 0.28]{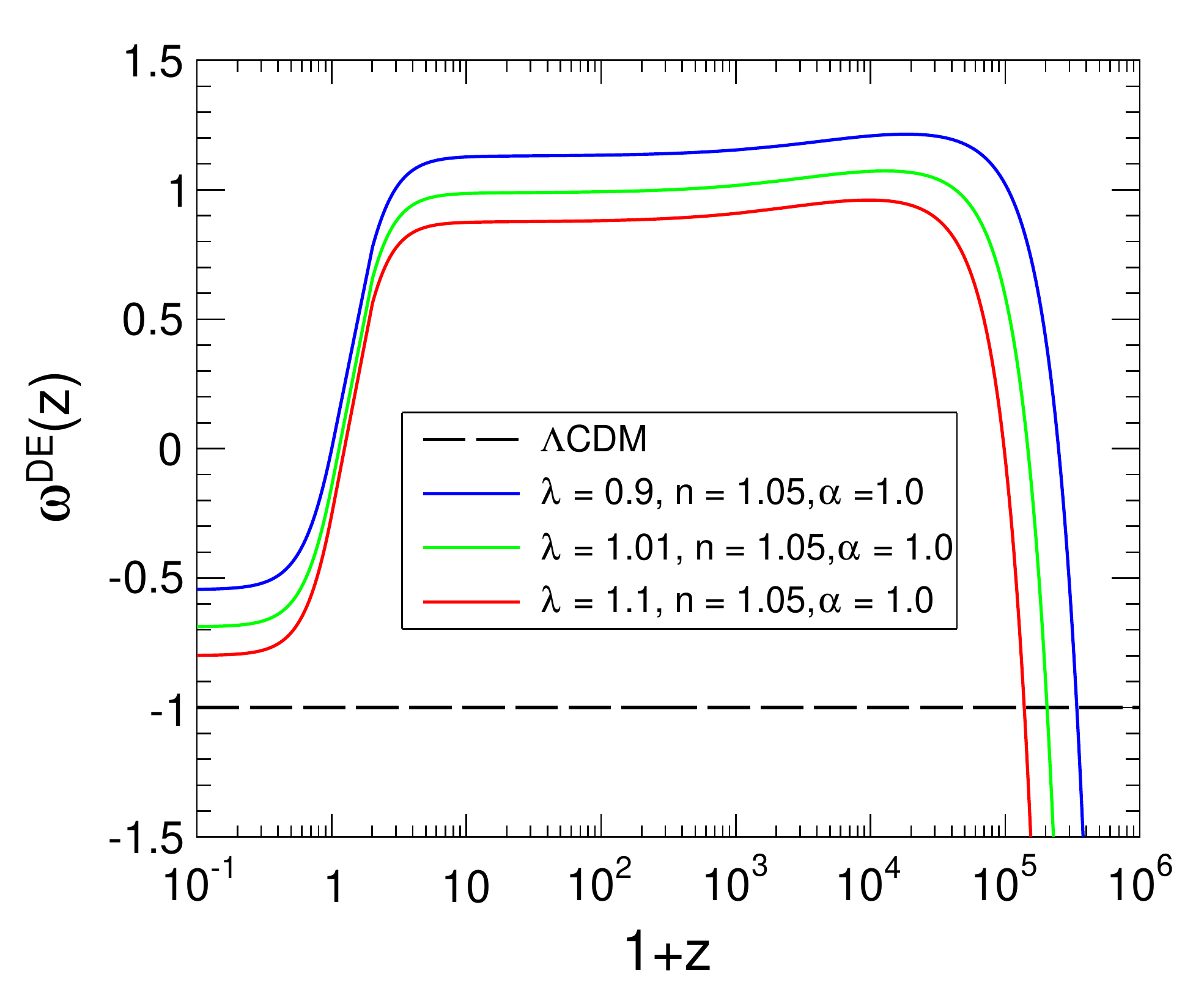}}
\vspace{-0.2cm}
\caption{DE EoS parameter $\omega^{DE}$ against cosmological redshift 
$z$ for different sets of values of model parameters as predicted by the power
law model in comparison with that of $\Lambda$CDM model.}
\label{fig2de}
\end{figure}

\begin{figure}[!htb]
\centerline{
  \includegraphics[scale = 0.28]{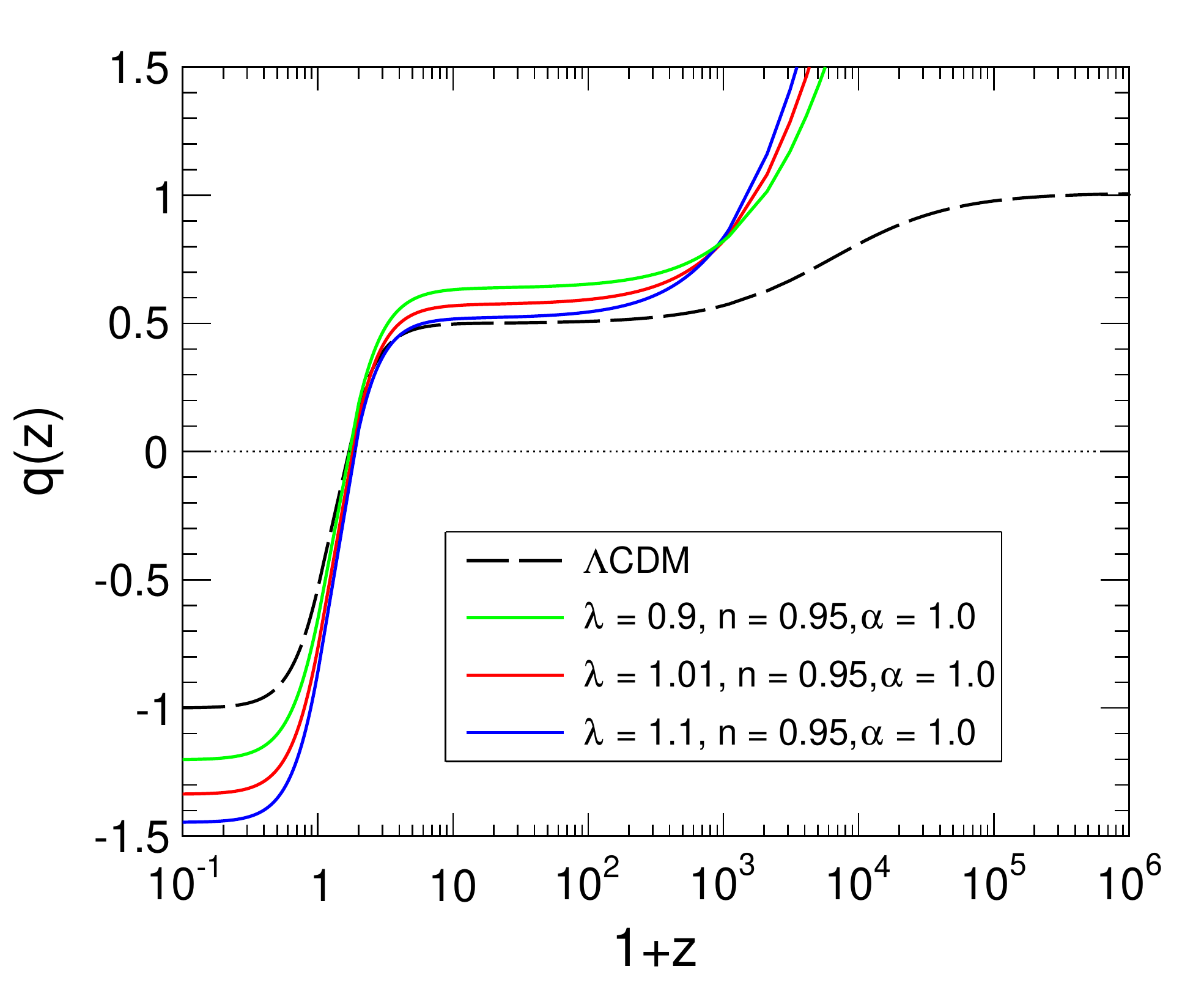}\hspace{0.25cm}
  \includegraphics[scale = 0.28]{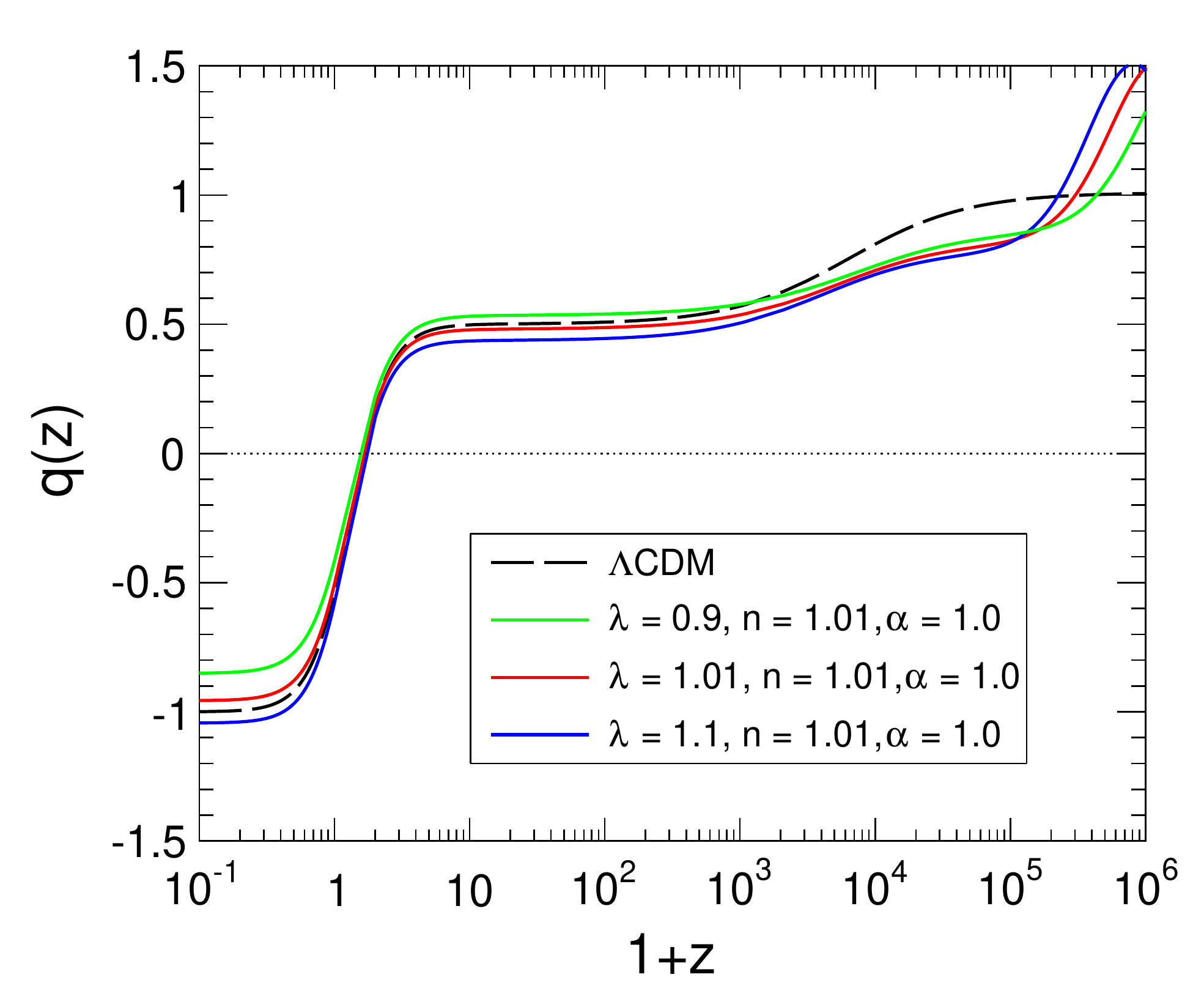}\hspace{0.25cm}
  \includegraphics[scale = 0.28]{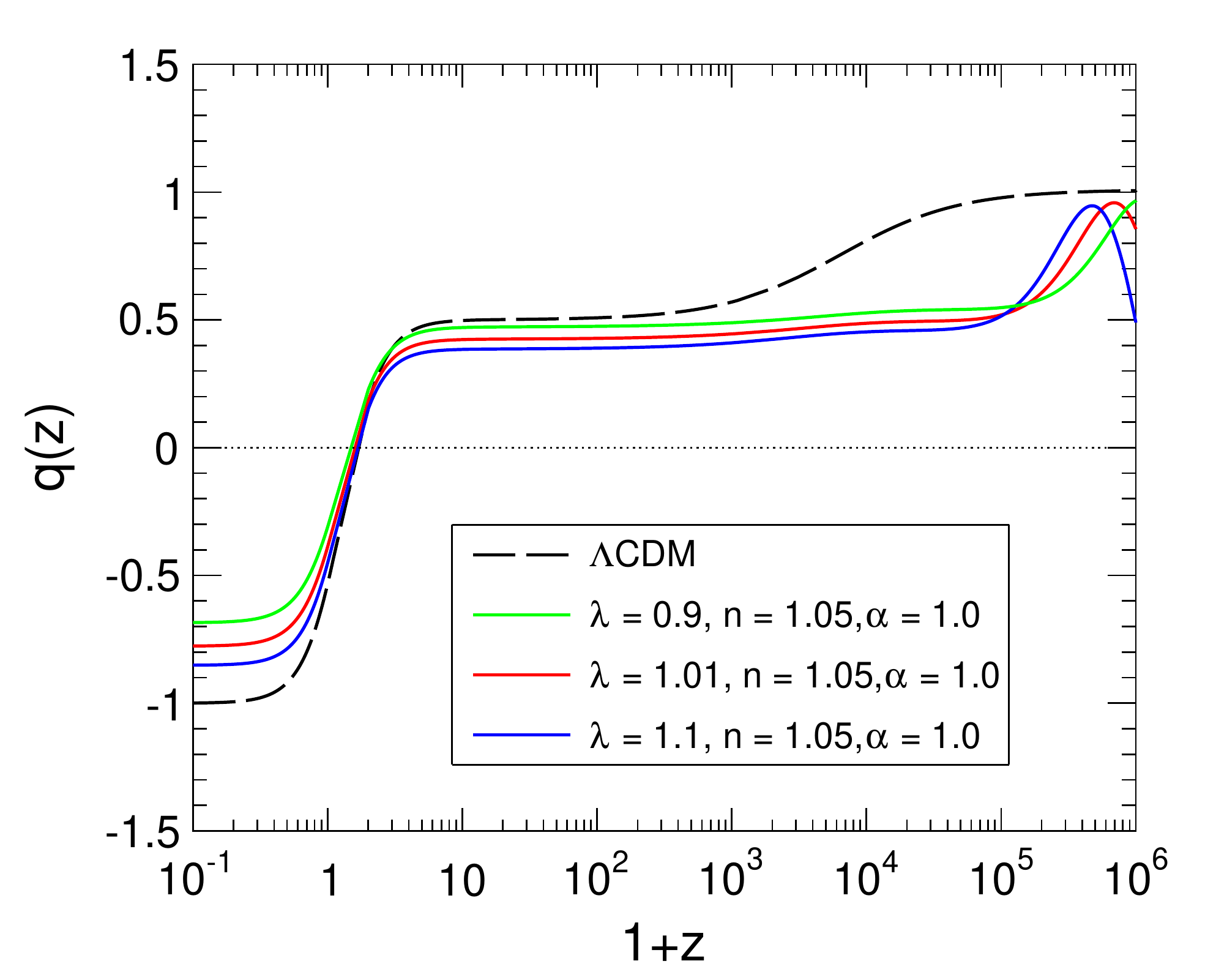}}
 \vspace{-0.2cm} 
\caption{Deceleration parameter $q(z)$ against cosmological redshift $z$ for
different values of model parameters as predicted by the power law model in 
comparison with that of $\Lambda$CDM model.}
\label{fig2d}
\end{figure}

We have selected three specific sets of model parameters from their constrained
ranges as mentioned above to plot the effective EoS parameter $\omega^{eff}$ , 
DE EoS parameter $\omega^{DE}$ and deceleration parameter $q$ against 
cosmological redshift $z$ on the basis of the agreements of these cosmological
parameters with the corresponding parameters in the $\Lambda$CDM Universe. It 
is found that behaviour of these plots are similar for the same value of $n$ 
with different values of other two parameters, but the behaviour changes 
significantly for different values of $n$. Therefore we have added three plots 
for three different values of $n$ within its constrained range for 
$\omega^{eff}$, $\omega^{DE}$ and $q$ in Figs.~\ref{fig2c}, \ref{fig2de} and \ref{fig2d}
 respectively.

From the above graphical analysis of effective EoS parameter $\omega^{eff}$ 
with respect to the cosmological redshift $z$ (Fig.~\ref{fig2c}) we have 
observed that this parameter deviates from the $\Lambda$CDM case for higher 
$z$ values. The pattern and prominence of this deviation depends on the set 
of model parameters used, specially on the value of the parameter $n$. It is 
seen that for higher values of $n$ the deviation becomes prominent even from 
comparatively smaller values $z$. However, for all sets of constrained model 
parameters this deviation becomes significant for the very large values of 
$z$ or in the very early Universe. On the other hand the present value of 
$\omega^{eff}$ for all sets of model parameters agrees almost well with its 
$\Lambda$CDM value. From the plots of the deceleration parameter $q$ against 
$z$ (Fig.~\ref{fig2d}), one can see that this parameter deviate notably from 
the $\Lambda$CDM result both in the early as well in the near future of the
Universe. For this parameter such deviations are significantly higher for 
smaller values of the parameter $n < 1$. For this model the behaviour of
$\omega^{DE}$ (Fig.~\ref{fig2de}) is peculiar and interesting throughout 
the evolution of the Universe. Except for the case with $n=0.95$, other cases 
with $n>1$ within the constraint range, $\omega^{DE}$ remains positive for the 
values of $z <\; \sim 10^4$. This is also the situation to be in near feature 
except the case of $\lambda = 1.1$. For the values of $z>\; \sim 10^4$, 
$\omega^{DE}$ falls back to large negative values, indicating large negative 
pressure and hence rapid expansion of the Universe. In the case with $n=0.95$, 
$\omega^{DE}$ always remains positive for $z>0$. Here, in the near the future 
part $\omega^{DE}$ remains negative and appears different from the expectation of the $\Lambda$CDM model, however it agrees in the current phase ($z=0$). Thus 
according to the power law model it can be concluded that although in the 
current scenario the role of anisotropy is not so prominent, but in the early 
as well as in the future Universe it may have or may play some vital role in
the evolution of the Universe.

\section{Discussion and conclusions}\label{sec6}
In this study, we have considered a symmetric teleparallel theory equivalent 
to GR (STEGR), and its extension the modified $f(Q)$ gravity theory and probed it in a Universe with small anisotropy (in terms of the LRS-BI spacetime) to examine the evolution of various cosmological parameters along with the 
effective equation of state and hence tried to study the evolution of the 
Universe. Here, we have started from deriving the non metricity scalar and 
then deriving the field equations for the LRS-BI metric. Since LRS-BI is an 
anisotropic model, thus we have also derived the shear scalar for our 
calculations. From the field equations, we have finally derived the effective 
EoS, DE EoS and other cosmological parameters. To close the system in such anisotropic scenario, we have considered the physically relatable condition $\sigma^2\propto \theta^2$ which in turn yielded the relation $H_x = \lambda H_y$, $\lambda$ being a constant. We have finally derived 
all the field equations and other cosmological parameters in accordance with 
this assumption using the average scale factor in terms of redshift $z$.

We have considered two $f(Q)$ models in this study. In the first model, we have 
taken the form of $f(Q) = -(Q + 2\Lambda)$ in which $\Lambda$ is the 
cosmological constant. All important cosmological parameters have been derived
using this model and found that all these expressions contained the
parameter $\lambda$. To constraint the value of $\lambda$ we have used the 
observational data. For this purpose, at first we have plotted the expression 
of the Hubble parameter obtained for the model with the HKP and SVJO5 
\cite{simon_2005}, SJVKS10 \cite{stern_2010}, and GCH09 \cite{hui_2009} data 
in comparison with $\Lambda$CDM model for up to $z=2$. It 
is found that the parameter $\lambda$ shows a good agreement with data 
within the range 0.5 to 1.25 of its values with the required restriction 
$\lambda \ne 1$ for the considered anisotropy. We have further plotted the 
distance modulus for the obtained range of $\lambda$ values along with SCP 
Union2.1 data \cite{suzuki_2012} and found that the 
plots excellently agree with the observational data as well as the 
$\Lambda$CDM model. Based on these two plots, thus we have found that 
$0.5 \le \lambda \le 1.25$ with $\lambda \ne 1$ is a reliable range for 
the parameter $\lambda$ in this $f(Q)$ model. The effective EoS parameter 
$\omega^{eff}$, DE EoS parameter $\omega^{DE}$ and deceleration parameter 
$q$ are plotted using the values of the parameter 
$\lambda$ within the constrained range with respect to redshift up to $10^6$. 
We found that although the model agrees with the current scenario and the 
near past ($z \leq 10^2$), it deviates significantly from the $\Lambda$CDM 
model in the early Universe. These deviations are due to the contribution of 
$\Omega_{\sigma}$ parameter in the expressions. Thus, from our analysis of this 
considered model of $f(Q)$, we have found that anisotropy may have some 
contributions in the early Universe. It is to be noted that for $\lambda = 1$, 
i.e.~in the isotropic case, the expressions reduce to that of the
$\Lambda$CDM model.

The second model considered in this study is in the form of power law as 
$f(Q) = -\alpha Q^{n}$ in which $\alpha$ and $n$ are the two model parameters. 
We have derived all expressions of cosmological parameters for this model as
we did in our previous model and found that all the cosmological parameters 
depend on three model parameters $\lambda$, $\alpha$ and $n$ in this case. We 
have employed the same technique to constrain these model parameters as in the 
previous case, i.e.~using the observational data of Hubble parameter and 
distance modulus, and plot them with the results from our expressions of Hubble 
parameter and distance modulus along with the $\Lambda$CDM model prediction. 
It is found that the effective ranges of parameters $\lambda$, $\alpha$ and
$n$ should be $0.5\le \lambda \le 1.25$ with $\lambda \ne 1$, 
$0.75 \le \alpha \le 1.5$ and $0.95 \le n \le 1.05$ with $n\ne 1$ 
respectively. With these values of model parameters we have plotted the 
effective EoS parameter $\omega^{eff}$, DE EoS parameter $\omega^{DE}$ and 
deceleration parameter $q$ with respect to the cosmological redshift $z$ and 
found that behaviours of these cosmological parameters are similar for same 
value of $n$ with different values of other two parameters, but their 
behaviours change significantly for different values of $n$, especially at 
higher $z$ values. We observed that these cosmological parameters deviate 
from their $\Lambda$CDM model predictions at higher values of redshift. Thus,  
the power law model also confirms the role of anisotropy in the early 
Universe. Further, we have found that these deviations are more prominent 
for higher values of $n$ in the case of effective EoS as in this case even in 
the lower $z$ values deviations are substantial. $\omega^{DE}$ shows 
prominent deviations from the $\Lambda$CDM prediction almost for all values
of $n$ with all values of other parameters nearly along the whole range of $z$ 
of study with some peculiar behaviours at very early phases of the Universe
depending on the value of $n$. In the case of the deceleration parameter $q$, 
we have found that it diverges too early for $n = 0.95$ as compared to the 
other two values of $n$. Again, for $n = 1.05$, the plot shows downward trend 
at very high values of $z$. It needs to be mentioned that all these plots 
reduce to $\Lambda$CDM ones, when all the three model parameters 
i.e.~$\alpha$, $\lambda$ and $n$ become equal to unity. Thus this model 
gives a scenario of our Universe where anisotropy has the dominating 
effect in its early phases and some possible effect in the near future. 

Therefore from our work we have tested the $f(Q)$ theory in LRS-BI metric 
to understand that the anisotropic nature of the Universe by considering a 
specific relation between directional Hubble parameters and found the 
existence of anisotropy in the early Universe for two different models of 
$f(Q)$ even though the signature of anisotropy is absent in the current stage 
of the Universe. These findings could be confirmed by the cosmological data 
of the early Universe, which may be available in the future from the 
advanced telescopes such as Thirty Meter Telescope \cite{TMT}, Extremely Large 
Telescope \cite{ELT}, CTA \cite{CTA} etc. In this context more rigorous analysis
of the anaisotropic models of the Universe based on the STEGR would be 
interesting and necessary in near future.


\section*{Acknowledgments}

A.D.~is supported in part by the FRGS research grant (Grant No. FRGS/1/2021/STG06/UTAR/02/1). U.D.G.~is thankful to the Inter-University Centre for Astronomy 
and Astrophysics (IUCAA), Pune, India for the Visiting Associateship of the 
institute.


\end{document}